\documentclass[prb,aps,showpacs]{revtex4}

\begin{document}
\title{Magnetization and dimerization profiles of the cut two-leg spin ladder}
\author{P. Lecheminant}
\affiliation{Laboratoire de Physique Th\'eorique et
Mod\'elisation, CNRS ESA 8089, 
Universit\'e de Cergy-Pontoise, 5 Mail Gay-Lussac, Neuville sur Oise, 
95301 Cergy-Pontoise Cedex, France}
\author{E. Orignac} 
\affiliation{Laboratoire de Physique Th\'eorique, CNRS UMR 8549,
Ecole Normale Sup{\'e}rieure, 24 rue Lhomond, 75231 Paris Cedex 05, France}
\begin{abstract} 
The physical properties of the edge states of the cut two-leg spin
ladder are investigated by means of the bosonization approach. By
carefully treating boundary conditions, we derive the existence of
spin-1/2 edge states in the spin ladder with a ferromagnetic rung
exchange and for the open spin-1 Heisenberg chain.  In contrast, such
states are absent in the antiferromagnetic rung coupling case.  The
approach, based on a mapping onto decoupled semi-infinite off-critical
Ising models, allows us to compute several physical quantities of
interest. In particular, we determine the magnetization and
dimerization profiles of the cut two-leg spin ladder and of the open
biquadratic spin-1 chain in the vicinity of the SU(2)$_2$ WZNW
critical point.
\end{abstract} 
\pacs{75.10.Jm 75.30.Hx 75.50.Ee 71.10.Pm}
\maketitle

\section{Introduction}
The infuence of impurities and imperfections on 
the behavior of low-dimensional strongly correlated
systems has attracted considerable attention in recent years.
The introduction of static non-magnetic impurities, like Zn or Li,
at the location of the magnetic ions 
is a sensitive probe of the correlations that develop
in these magnetic systems.
A particularly striking exemple is the observation of 
fractional spin-1/2 edge states in the Haldane
gap\cite{haldane_gap} spin-1 compound NENP
cut by non-magnetic impurities.\cite{glarum}
These spin-1/2 degrees of freedom are associated with 
static staggered moments 
close to the chain ends which
are revealed unambiguously in 
the NMR profile of the Mg-doped Y$_2$BaNiO$_5$. \cite{tedoldi} 
This effect can be viewed as a local restoration of
antiferromagnetism by impurities. 
In fact, a transition to an antiferromagnetic state induced by the
local moments has been observed recently in the Haldane gap compound
$\mathrm{PbNi_2V_2O_8}$.\cite{uchiyama_s1_imp} 
Such a local enhancement of antiferromagnetism
induced by non-magnetic impurities
is a rather general phenomenon
in gapful quasi-one dimensional systems.\cite{martins} 
The  spin ladder
material $\mathrm{SrCu_2O_3}$ lightly doped with
Zn impurities exhibits a Curie-like behavior at low 
temperature which has been explained by
the unpaired free spins in the 
vicinity of the impurity.\cite{azuma_zinc_doping}
Further, it has
been shown by NMR measurements 
that a staggered
magnetization is induced along the leg by very small (0.25\%)
concentrations of impurities.\cite{fujiwara_srcuo_imp,ohsugi_srcuo_imp} 
At low temperature, the induced
moments get frozen, leading to N\'eel
order.\cite{azuma_zinc_doping,ohsugi_srcuo_imp,larkin_2leg_imp}
Similar  effects have been observed in the ladder compound
$\mathrm{Cu_2(C_5H_{12}N_2)_2Cl_4}$
doped with Zn impurities \cite{deguchi_cuhpcl_imp} and  
in the spin-Peierls gap  material
CuGeO$_3$.\cite{hase_cugeo3_zn}

A simple explanation of the existence of free spin-1/2
moments at the ends of a broken spin-1 chain 
can be obtained from 
the Valence Bond Solid (VBS) model\cite{affleck_klt_short}
where each S=1 spin is viewed as two S=1/2 spins in
the symmetric triplet state. 
In this model, 
the singlet ground state of
a chain with periodic boundary
is described by two
valence bonds originating from each site to form singlets
with adjacent neighbors. If the chain is broken (i.e. open
boundary conditions are considered), unpaired bonds are left
at each end of the chain resulting in two free S=1/2 objects
at the boundaries and a fourfold ground state degeneracy.
This VBS picture provides a good and intuitive description
of the ground state of the spin-1 chain.
In particular, 
the exact diagonalization\cite{kennedy} of finite open samples
with even number of sites
has shown that the ground state is a singlet and 
the existence of an exponentially low-lying triplet state in the Haldane gap.
This leads to a fourfold ground state degeneracy in the thermodynamic
limit.
Such degeneracy can also be interpreted as the consequence of
a spontaneously broken hidden Z$_2 \times$ Z$_2$
symmetry \cite{kennedy_z2z2_haldane} associated with the formation of
the Haldane gap. A string order parameter 
has been introduced to reveal this hidden
symmetry.\cite{nijs_dof,kim}
More generally, it is expected within the VBS and 
non-linear sigma model approaches
that the integer spin-S Heisenberg 
chain has spin-S/2 chain-end
excitations.\cite{ng_schwinger}
   
The physical properties of the S=1/2 chain-boundary excitations 
in the open spin-1 chain have been investigated in details in a  
quantum Monte Carlo (QMC) study\cite{miyashita} and by means of the 
density-matrix renormalization
group (DMRG) approach.\cite{white_dmrg_letter,sorensen_dmrg,qinng,batista,legeza_spins}
Recently, the magnetization profiles at finite temperatures and fields
have
been determined using continuous time QMC techniques 
to reconstruct the experimentally 
measured NMR spectrum of the Mg-doped Y$_2$BaNiO$_5$.\cite{alet_doped_s=1}
The properties of the edge states of a more general model,
the bilinear biquadratic 
spin-1 Heisenberg chain, defined by the Hamiltonian:
\begin{equation}
{\cal H}=J\sum_i \left[ \mathbf{S}_i\cdot\mathbf{S}_{i+1} + \beta
(\mathbf{S}_i\cdot\mathbf{S}_{i+1})^2\right],
\label{biquaham}
\end{equation}
have also been considered.\cite{polizzi_boundary_s=1}
For $\beta=1/3$, (the so-called AKLT point)  the 
VBS state turns out to be the exact ground
state of the bilinear-biquadratic
model.\cite{affleck_klt_short}  
For $\beta=-1$, the Hamiltonian (\ref{biquaham}) has
a critical point separating the
Haldane phase ($-1 <\beta < 1$) from a dimerized phase ($\beta<-1$). 
At this ($\beta=-1$) 
critical point, the model is
integrable\cite{takhtajan_spin_s} and belongs to the
SU(2)$_2$ Wess-Zumino-Novikov-Witten (WZNW) 
universality class.\cite{affleck_strongcoupl}
The authors of Ref. \onlinecite{polizzi_boundary_s=1} 
have established that
the edge states are present through the whole Haldane phase and
disappear as soon as the $\beta=-1$ critical point is reached.

In this paper, we shall investigate the physical properties 
of the S=1/2 chain-end excitations of 
the semi-infinite (or cut) two-leg spin ladder and of the 
open spin-1 chain
by means of the bosonization
method.\cite{gogolin_book} 
In the strong ferromagnetic rung limit, this two-leg ladder model
is equivalent to the open spin-1 Heisenberg chain with the two spins
on the rung forming an effective S=1 local moment.
Since this strong coupling limit is smoothly connected to
the weak coupling one,\cite{hida_2ch,watanabe_ladder_obc,white}
the approach provides a simple way to extract
the low-energy properties
of the open spin-1 chain.
In this respect, it gives an alternative derivation
of the existence of the spin-1/2 edge states 
predicted by the VBS theory\cite{affleck_klt_short}
and the Schwinger-boson mean-field analysis.\cite{ng_schwinger}
Futhermore, we shall be able to calculate explicitly 
the physical properties of the open spin-1 chain
such as the magnetization profile or the NMR relaxation rate.
To this end, the mapping\cite{shelton_spin_ladders} 
of the low-energy Hamiltonian of a weakly coupled two-leg ladder
onto off-critical two-dimensional Ising models will be exploited
to derive the chain-boundary excitations
as it has been done 
in the study of disordered spin-1/2 ladders.\cite{gogolin_disordered_ladder}
By paying a careful attention on
the boundary conditions on the fields that occur in the continuum
limit, 
the staggered magnetization profiles of the model can be determined
using exact results of semi-infinite one-dimensional 
quantum Ising model.
The results strongly depend on the sign of the interchain coupling
and for an antiferromagnetic rung exchange, no magnetic chain-end
excitations are found. 
However, a weak dimerization, induced by 
the presence of the boundary,\cite{ng_schwinger,qinng,tsai}
exists for all signs of the interchain interaction and
can be computed by this mapping onto semi-infinite Ising models.
Finally, the influence of a strong external magnetic field
fixing the spins at the edge can be investigated by a
similar approach. 

The rest of the paper is organized as follows.
In Section \ref{sec:deriv-effect-low}, the low-energy Hamiltonian of the cut two-leg
spin ladder is mapped onto an O(3)$\times$ Z$_2$ symmetric
theory of four massive Majorana fermions with suitable boundary
conditions. The nature of the edge states that occur in the 
problem is then
discussed in Section \ref{sec:s=12-chain-boundary} where the uniform
magnetization profile and the NMR relaxation rate are computed
for a ferromagnetic interchain interaction.
Section \ref{sec:stagg-profile} presents the 
calculation of the staggered magnetization
and dimerization profiles of the model by exploiting the mapping
onto semi-infinite off-critical quantum Ising models. 
The effect of a strong applied magnetic field fixing
the spins at the boundary is investigated in Section
\ref{sec:strong-extern-field}. 
Finally, our concluding remarks are presented in Section
\ref{sec:concluding-remarks} 
and the paper is supplied with four Appendices which
provide some technical details used in this work.

\section{Derivation of the effective low-energy
Hamiltonian}\label{sec:deriv-effect-low} 

In this section, we apply the bosonization approach
to the semi-infinite two-leg spin ladder described by
the Hamiltonian
\begin{eqnarray}\label{eq:cut_ladder}
{\cal H} =\sum_{n=0}^\infty \left[ J_\parallel \sum_{p=1,2} {\bf S}_{n,p}\cdot
{\bf  S}_{n+1,p} + J_\perp {\bf S}_{n,1}\cdot {\bf S}_{n,2} \right],
\end{eqnarray}
where ${\bf S}_{n,p}$ is a spin-1/2 operator at site $n$ 
on chain $p$ ($p=1,2$) and we consider
an antiferromagnetic inchain interaction $J_\parallel >0$.
The bosonization method will be applied
to the Hamiltonian (\ref{eq:cut_ladder})
in the regime $|J_\perp| \ll J_\parallel$ and
with a suitable redefinition of the
effective coupling constants it captures the physical properties 
of the model for
arbitrary $J_\perp$ since there is a continuity between
the weak and strong coupling limits in the
two-leg spin ladder.\cite{hida_2ch,watanabe_ladder_obc,white}
In particular, local S=1 spins are formed in each rung
of the ladder in the strong ferromagnetic interchain coupling limit
($J_{\perp} < 0$) so that the approach provides, in turn, a way to
investigate the physical properties 
of the broken spin-1 Heisenberg chain.

\subsection{Bosonization of the open two-leg spin ladder}
Let us first consider the decoupling limit ($J_\perp=0$) where the system
reduces to two independent spin-1/2 Heisenberg chains with open boundary 
conditions. The low-energy properties of the latter
model can be still determined by means of the 
bosonization method.\cite{eggert_openchains,wong,fabrizio_open_electron_gas,ng_qin,mattson,hikihara}
As it is described in the Appendix A, starting from the underlying 
Hubbard model,
the bosonized Hamiltonian for the open spin-1/2 Heisenberg
chain with index $p=1,2$ reads as follows neglecting the 
marginally irrelevant term:
\begin{equation}
{\cal H}_p^0 = 
\frac{v}{2\pi} \int_0^\infty dx \left[ (\pi \Pi_p)^2 +(\partial_x
\Phi_p)^2\right], 
\end{equation}
where $v$ is the spin velocity and $\Pi_p$ is the momentum
operator conjugate to the bosonic field $\Phi_p$.
The boundary condition on the fields  $\Phi_p$ reads
\begin{equation}
\Phi_p\left(0\right)=0,
\label{inibondfield}
\end{equation}
which corresponds to a Dirichlet boundary condition. 
In the continuum limit, the effective spin density ${\bf S}_p(x)$
separates into an uniform and staggered parts:
\begin{equation}
{\bf S}_p \left(x\right) = {\bf J}_{p R} \left(x\right) + 
{\bf J}_{p L} \left(x\right) + \left(-1\right)^{x/a} {\bf n}_p \left(x\right),
\label{inispindens}
\end{equation}
$a$ being the lattice spacing.
As shown in the Appendix A, the bosonized 
description for the uniform spin density is given by 
\begin{eqnarray}
J_{p R,L}^{z} &=& - \frac{1}{2\pi\sqrt{2}} \ \partial_x \Phi_{p R,L} \nonumber \\
J_{p R}^{\dagger} &=&  \frac{e^{i \sqrt{2} \Phi_{p R}}}{2\pi a}
\nonumber \\
J_{p L}^{\dagger} &=& \frac{e^{-i \sqrt{2} \Phi_{p L}}}{2\pi a},
\label{inspinuniboso}
\end{eqnarray}
$\Phi_{p R,L}$ being the chiral components of the bosonic
field $\Phi_p$: $\Phi_p = (\Phi_{p R} + \Phi_{p L})/2$.
The staggered part of the 
spin density (\ref{inispindens}) can be expressed in terms of $\Phi_p$
and its dual field $\Theta_p$:
\begin{equation}
{\bf n}_p = \frac{\lambda}{\pi a} \left[
\cos\left(\sqrt{2} \ \Theta_p\right), 
-\sin\left(\sqrt{2}\ \Theta_p\right),
-\sin\left(\sqrt{2}\ \Phi_p\right) \right],
\label{inspinstagboso}
\end{equation}
where $\lambda$ is a constant stemming from the underlying charge degrees of
freedom that have been integrated out. 

In the weak coupling regime $|J_{\perp}| \ll J_{\parallel}$, 
the continuum limit of the Hamiltonian (\ref{eq:cut_ladder})
can then be derived using all these results. 
To this end, we introduce the 
symmetric and antisymmetric combinations of the 
bosonic fields:
\begin{equation}
\Phi_{\pm R,L} = \frac{\Phi_{1 R,L} \pm \Phi_{2 R,L}}{\sqrt{2}},
\label{symantisym}
\end{equation}
so that the leading part of the Hamiltonian (\ref{eq:cut_ladder}) that
imposes the strong coupling behavior of the system
decomposes into two commuting 
parts ${\cal H}_{\pm}$:\cite{schulz_spins,shelton_spin_ladders}
\begin{equation} 
{\cal H} \simeq {\cal H}_{+} + {\cal H}_{-} \  \left[{\cal H}_{+}, {\cal H}_{-}
\right] = 0,
\label{hcont}
\end{equation}  
with the following bosonized expressions
\begin{eqnarray}\label{eq:ladder_bosonized}
& & {\cal H}_+ = \frac{v}{2\pi} \int_0^\infty dx \;
\lbrack (\pi \Pi_+)^2+(\partial_x
\Phi_+)^2\rbrack 
-\frac{J_{\perp}\lambda^2}{2\pi^2 a}\int_0^\infty dx \; \cos 2 \Phi_+ \nonumber\\
\label{ha}
& & {\cal H}_- = \frac{v}{2\pi} \int_0^\infty dx  \;
\lbrack (\pi \Pi_-)^2+(\partial_x
\Phi_-)^2\rbrack +\frac{J_{\perp}\lambda^2}{2\pi^2 a}\int_0^\infty dx \; \cos2\Phi_-
+\frac{J_{\perp}\lambda^2}{\pi^2 a}\int_0^\infty dx \; \cos2\Theta_-,
\end{eqnarray}
where the boundary conditions on the 
bosonic fields are of Dirichlet type:
\begin{equation}
\Phi_{\pm}\left(0\right) = 0.
\label{bounphipm}
\end{equation}
In this derivation of the low-energy theory, one should note that we 
have only 
taken into account the most relevant perturbation that appears
in the continuum limit of the spin ladder. In particular,
we have discarded the marginal contribution that stems
from the uniform pieces of the spin densities (\ref{inispindens}).
We shall later comment on its main effect when the Hamiltonian 
(\ref{hcont}) will be refermionized.

\subsection{Refermionization}

The next step of the approach is to observe that
the scaling dimension of the interacting part in 
${\cal H}_{\pm}$ is equal to one. 
The bosonic fields are precisely at the free-fermion point
where the cosine terms in Eq. (\ref{eq:ladder_bosonized})
can be expressed in terms of massive fermions.\cite{shelton_spin_ladders}
To this end, we first introduce the left and right bosonic fields
corresponding to $\Phi_{\pm}$:
\begin{eqnarray}\label{eq:phiR_phiL}
\Phi_{\pm} &=& \frac 1 2 \left( \Phi_{\pm L} +
\Phi_{\pm R} \right) \nonumber \\
\Theta_{\pm} &=& \frac 1 2 \left( 
\Phi_{\pm L} - \Phi_{\pm R} \right).
\end{eqnarray}
These chiral fields are no longer independent due to
the existence of the boundary condition (\ref{bounphipm})
and one has:
\begin{equation}
\Phi_{\pm R} \left(0\right) = - \Phi_{\pm L} \left(0\right).
\label{bonphipmchir}
\end{equation}
The refermionization of the Hamiltonian (\ref{eq:ladder_bosonized})
can then be obtained through the bosonization formula
\begin{eqnarray}\label{fermi_fields}
\psi_{\pm R} &=& \frac{\kappa_{\pm} \; e^{-i\Phi_{\pm R}}}{\sqrt{2\pi a}} 
\nonumber \\
\psi_{\pm L} &=& \frac{\kappa_{\pm} \; e^{i\Phi_{\pm L}}}{\sqrt{2\pi a}},
\end{eqnarray}
where $\kappa_{\pm}$ are Klein factors that
obey the anticommutation
relation $\{\kappa_{+}, \kappa_{-} \} = 0$
to ensure the anticommutation between the
fermion fields with different channel index $\pm$.
The anticommutation between $\psi_{\pm R}$ and $\psi_{\pm L}$
results from
$[\Phi_{\pm R}(x), \Phi_{\pm L}(y)] = - i \pi$ which
stems from the 
Dirichlet boundary condition (\ref{bounphipm}) 
as described in the Appendix A. 
The boundary conditions on
the fermionic fields can be deduced 
from Eq. (\ref{bonphipmchir}) 
\begin{equation}\label{eq:continuation_fermi}
\psi_{\pm R} \left(0 \right) =  \psi_{\pm L} \left(0 \right).
\end{equation}
The cosine terms of Eq. (\ref{eq:ladder_bosonized})
can then be refermionized using the identification 
(\ref{fermi_fields}) as well as 
the commutation 
relation $[\Phi_{\pm R}(x), \Phi_{\pm L}(x)] = - i \pi, x > 0$:
\begin{eqnarray}
\cos 2 \Phi_{\pm} &=& - i \pi a \left(\psi^\dagger_{\pm R} \psi_{\pm L} -
\psi^\dagger_{\pm L} \psi_{\pm R}\right) \nonumber \\
\cos 2 \Theta_{\pm} &=& \; \; \; i\pi a \left(\psi^\dagger_{\pm R}
\psi^\dagger_{\pm L} -  \psi_{\pm L} \psi_{\pm R} \right).
\label{refercos}
\end{eqnarray}
The Hamiltonians ${\cal H}_{\pm}$ of Eq. (\ref{eq:ladder_bosonized})
can thus be expressed in terms of the fermion fields
\begin{eqnarray}
{\cal H}_+ &=& -i v \int_0^\infty dx \; 
\left(\psi_{+ R}^\dagger \partial_x \psi_{+ R}
-\psi_{+ L}^\dagger \partial_x \psi_{+ L}\right) 
+ \frac{i J_\perp \lambda^2}{2\pi}
\int_0^\infty dx \; \left(\psi_{+ R}^\dagger \psi_{+ L} - \psi_{+ L}^\dagger
\psi_{+ R}\right) \nonumber \\
{\cal H}_- &=& -i v \int_0^\infty dx \; \left(\psi_{-R}^\dagger 
\partial_x \psi_{- R}
-\psi_{- L}^\dagger \partial_x \psi_{- L}\right) 
- \frac{i J_\perp \lambda^2}{2\pi}
\int_0^\infty dx \; \left(\psi_{- R}^\dagger \psi_{- L} - \psi_{- L}^\dagger
\psi_{- R}\right) 
\nonumber \\
&+& \frac{i J_\perp \lambda^2}{\pi} \int_0^\infty dx \left(\psi_{- R}^\dagger
\psi_{- L}^\dagger - \psi_{- L} \psi_{- R}\right), 
\label{ham2legdirac}
\end{eqnarray}
with the boundary conditions (\ref{eq:continuation_fermi}) for the 
fermion fields.

At this point, it is convenient to
introduce four Majorana (real) fermions from the 
Dirac ones (\ref{fermi_fields})
\begin{eqnarray}
\psi_{+ R,L} &=& \frac{\xi_{R,L}^2 + i\xi_{R,L}^1}{\sqrt{2}} \nonumber \\
\psi_{- R,L} &=& \frac{\xi_{R,L}^3 + i\xi_{R,L}^0}{\sqrt{2}}.
\label{majorana}
\end{eqnarray}
This identification together with 
the correspondences (\ref{inspinuniboso},\ref{fermi_fields})
enable us to derive the refermionization 
of the uniform part of the spin densities (\ref{inispindens}) 
\begin{eqnarray}
I^{a}_{R,L} &=& J^{a}_{1 R,L} + J^{a}_{2 R,L} = -\frac{i}{2} 
\epsilon^{a b c} \xi_{R,L}^{b} \xi_{R,L}^{c} \nonumber \\
K^{a}_{R,L} &=& J^{a}_{1 R,L} - J^{a}_{2 R,L} = \; i \xi_{R,L}^{a} 
\xi_{R,L}^{0},
\label{currmajo}
\end{eqnarray}   
where we have fixed the product $\kappa_{+} \kappa_{-}$ of the Klein factors 
that appear in Eq. (\ref{fermi_fields}) to $i$ to
obtain Eq. (\ref{currmajo}). In fact, this identification
(\ref{currmajo}) is nothing but the faithful representation of two
independent SU(2)$_1$ Kac-Moody currents ${\bf J}_{p R,L}, p=1,2$
in terms of four Majorana
fermions.\cite{zamolodchikov_fateev,allen,gogolin_book}

With the above results, the Hamiltonian (\ref{hcont}) can be rewritten
with the four Majorana fermions
and  be separated into
two commuting (triplet and singlet) pieces
\begin{equation}
{\cal H} = {\cal H}_t + {\cal H}_s,
\label{septrisin}
\end{equation}
with
\begin{eqnarray}\label{eq:majorana_hamiltonian}
{\cal H}_t &=& -\frac{i v}{2} \int_0^\infty dx
\; \sum_{a=1}^{3} \left(\xi_R^a
\partial_x \xi_R^a -  \xi_L^a  \partial_x \xi_L^a\right) +
\frac{i J_\perp \lambda^2}{2\pi} \int_0^\infty dx \sum_{a=1}^{3}
\; \xi_R^a \xi_L^a \nonumber \\
{\cal H}_s &=& -\frac{i v}{2} \int_0^\infty dx
\; \left(\xi_R^0
\partial_x \xi_R^0 -  \xi_L^0  \partial_x \xi_L^0\right) 
-\frac{3 i J_\perp \lambda^2}{2\pi} \int_0^\infty dx 
\; \xi_R^0 \xi_L^0 . 
\end{eqnarray}
The boundary conditions on the Majorana fermions are
 obtained from the constraint (\ref{eq:continuation_fermi})
and the definition (\ref{majorana}) 
\begin{equation}\label{eq:majorana_bc}
\xi_R^a \left(0\right) = \xi_L^a \left(0\right), \; a=0,\ldots,3 .
\end{equation}
Moreover, the marginal interchain perturbation that we 
have so far neglected can be expressed in terms of the 
Majorana fermions using the correspondence (\ref{currmajo}).
As shown in Ref. \onlinecite{shelton_spin_ladders}, the resulting contribution
leads to a velocity anisotropy and a mass-renormalization 
in the singlet and triplet sectors so that the low-energy
Hamiltonian (\ref{septrisin}) takes now the following form
\begin{eqnarray}\label{eq:majorana_hamiltonianrenor}
{\cal H}_t &=& -\frac{i v_t}{2} \int_0^\infty dx
\; \sum_{a=1}^{3} \left(\xi_R^a
\partial_x \xi_R^a -  \xi_L^a  \partial_x \xi_L^a\right) -
i m_t\int_0^\infty dx \sum_{a=1}^{3}
\; \xi_R^a \xi_L^a \nonumber \\
{\cal H}_s &=& -\frac{i v_s}{2} \int_0^\infty dx
\; \left(\xi_R^0
\partial_x \xi_R^0 -  \xi_L^0  \partial_x \xi_L^0\right)
-i m_s \int_0^\infty dx
\; \xi_R^0 \xi_L^0, 
\end{eqnarray}                    
where $m_t > 0$ and $m_s < 0$ (respectively $m_t < 0$ and $m_s > 0$)
for a ferromagnetic (respectively antiferromagnetic) 
interchain coupling and in particular
in the weak coupling case $|J_{\perp}| \ll J_{\parallel}$ 
one has the identification
from Eq. (\ref{eq:majorana_hamiltonian}): $m_t = - J_{\perp}\lambda^2/2\pi$
and $m_s = 3 J_{\perp}\lambda^2/2\pi$.

We thus observe that in the low-energy limit the
initial Hamiltonian (\ref{eq:cut_ladder}) of the cut two-leg spin
ladder is mapped onto a model of four free massive Majorana
fermions with the boundary conditions (\ref{eq:majorana_bc}).
In the strong ferromagnetic rung limit $-J_{\perp} \gg J_{\parallel}$, 
the singlet excitation described by the Majorana 
fermion $\xi_{R,L}^{0}$ are frozen ($|m_s| \to \infty$) so that the low-energy
properties of the model are governed by the triplet
magnetic excitations corresponding 
to the fields $\xi^{a}_{R,L}, a=1,2,3$. In this strong ferromagnetic rung
limit, we expect the system to be equivalent to a broken spin-1 chain. 
Indeed, it was shown by Tsvelik\cite{tsvelik_field}
by perturbing around  the SU(2)$_2$ WZNW 
critical point\cite{takhtajan_spin_s} 
of the biquadratic spin-1 chain, which is described by 
three massless Majorana fermions, that the low-energy
properties of a gapped spin-1 chain could be obtained from a
triplet of massive Majorana fermions. Furthermore, it can be seen easily that
the boundary conditions (\ref{eq:majorana_bc}) imply that the 
SU(2)$_2$ currents obey
$I_R^a(0)=I_L^a(0)$ which means that there is no spin current flowing
across the boundary. Therefore,  an open biquadratic spin-1
chain is described by a triplet of massive Majorana fermions: 
\begin{eqnarray}\label{eq:majorana_h_spin1}
{\cal H}_t &=& -\frac{i v_t}{2} \int_0^\infty dx
\; \sum_{a=1}^{3} \left(\xi_R^a
\partial_x \xi_R^a -  \xi_L^a  \partial_x \xi_L^a\right) -
i m_t\int_0^\infty dx \sum_{a=1}^{3}
\; \xi_R^a \xi_L^a ,
\end{eqnarray}
\noindent with the boundary conditions (\ref{eq:majorana_bc}) and
the Haldane (respectively dimerized) phase is characterized
by a positive (respectively negative) triplet mass $m_t$.

\section{S=1/2 chain-boundary excitations}\label{sec:s=12-chain-boundary}
In this section, the nature of the edge states of
the cut two-leg spin ladder and open spin-1 chain 
are investigated using the
low-energy description (\ref{eq:majorana_hamiltonianrenor}) of the model 
in terms of four Majorana fermions with boundary conditions.
In particular,  physical quantities such as the uniform component
of the magnetization profile and the NMR relaxation rate
will be computed within this approach.
The calculation of the staggered magnetization near the edge
will be presented in the next section since it involves 
quantities that are nonlocal in terms of the Majorana fermions.

\subsection{Localized Majorana fermion state}
The special structure of the low-energy Hamiltonian 
(\ref{eq:majorana_hamiltonianrenor}) together with the
constraint (\ref{eq:majorana_bc}) lead us to consider
a single massive Majorana fermion Hamiltonian
of the form
\begin{equation}
{\cal H}_{\textrm{toy}} = 
\frac{1}{2} \int_{0}^{\infty} dx \;
\Psi \left(x\right)^{T} \left( -i v \sigma_3 \partial_x + 
m \sigma_2 \right) \Psi \left(x\right),
\label{hamtoy}
\end{equation}
where $\sigma_i$ are the usual Pauli matrices and 
$\Psi \left(x\right)$ is a Majorana 2-spinor that writes 
\begin{equation}
\Psi \left(x\right) =
\left(\begin{array}{c} \xi_R\left(x\right) \\ 
\xi_L\left(x\right) \end{array} \right),
\end{equation}
with boundary condition $\xi_R(0) = \xi_L(0)$. 
The Hamiltonian (\ref{hamtoy}) is exactly-solvable
being quadratic in terms of the fermions and the 
resulting eigenvectors read as follows in the Heisenberg
representation
\begin{equation}\label{eq:eigen_decomposition}
\Psi\left(x,t\right) = \frac 1 {\sqrt{2L}} \sum_{k>0} 
\left\{\xi_{k} \left(\begin{array}{c} \cos
\left(kx+\theta_k\right) + i \sin\left(kx\right) \\  \cos
\left(kx+\theta_k\right) - i \sin\left(kx\right)\end{array}\right) 
e^{-i \epsilon_k t}
+ H. c. \right\}
 +  \sqrt{\frac m v} \left( \begin{array}{c} 1 \\ 1 \end{array}
\right) e^{-mx/v} \; \theta\left(m\right) \eta, 
\end{equation}
where $\xi_k$ is a fermion annihilation operator
with $k=\pi n/L$, $\eta$ is a zero mode Majorana
fermion, and $\theta$ is the Heaviside step function.
In Eq. (\ref{eq:eigen_decomposition}), $\epsilon_k$
denotes the energy dispersion of the model: 
\begin{equation}
\epsilon_k =\sqrt{v^2 k^2 +m^2},
\label{dispersion}
\end{equation}
and $\theta_k$ is given by
\begin{eqnarray}\label{eq:bogoliubov_rotation}
\cos \theta_k &=& \frac{vk}{\epsilon_k} \nonumber\\
\sin \theta_k &=& \frac{m}{\epsilon_k}.
\end{eqnarray}
For a positive mass $m$, one observes from the 
decomposition (\ref{eq:eigen_decomposition}) the
existence of an exponentially localized state
with zero energy inside the gap.
Such localized Majorana fermionic states have 
already been discussed in several different contexts such 
as  the holon edge state in 
an attractive one-dimensional electron gas,\cite{fabrizio_open_electron_gas,lopatin} the random mass Majorana fermion model,\cite{balents_random_dirac,shelton_disorder,gogolin_disordered_ladder}
and the problem of a magnetic impurity in 
a superconductor.\cite{gogolinsupra,bassi}  Finally 
it has been pointed out recently that such bound states
may find applications in quantum computation.\cite{kitaev,levitov_qbits}

>From this analysis of the toy model (\ref{hamtoy}), 
we deduce the decomposition of the triplet and singlet 
Majorana fields in the basis of the eigenvectors 
of the Hamiltonian (\ref{eq:majorana_hamiltonianrenor}) subject to the
boundary conditions (\ref{eq:majorana_bc}). 
For $J_\perp <0$ i.e. $m_t > 0$ and $m_s <0$, one obtains for the triplet
sector $a=1,2,3$ with obvious notations
\begin{equation}
\left(\begin{array}{c} \xi_R^a \\ \xi_L^a \end{array} \right)
\left(x,t\right) = \frac 1 {\sqrt{2L}} \sum_{k>0} 
\left\{\xi^{a}_k \left(\begin{array}{c} \cos
\left(kx+\theta_k^t\right) + i \sin\left(kx\right) \\  \cos
\left(kx+\theta_k^t\right) - i \sin\left(kx\right) 
\end{array} \right) e^{-i \epsilon^t_k t} + H.c.
\right\}
 + \sqrt{\frac{m_t}{v_t}} \left(\begin{array}{c} 1 \\ 1 \end{array}
\right) e^{-m_t x/v_t} \eta^a, 
\label{decomptriplet}
\end{equation}
whereas the decomposition for the singlet excitations with $m_s < 0$
reads
\begin{equation}
\left(\begin{array}{c} \xi_R^0 \\ \xi_L^0 \end{array} \right)
\left(x,t\right) = \frac 1 {\sqrt{2L}} \sum_{k>0}
\left\{\xi^{0}_k \left(\begin{array}{c} \cos
\left(kx+\theta_k^s\right) + i \sin\left(kx\right) \\  \cos
\left(kx+\theta_k^s\right) - i \sin\left(kx\right)
\end{array} \right) e^{-i \epsilon^s_k t} + H.c.
\right\}. 
\label{decompsinglet}
\end{equation}                     
Therefore, the localized Majorana zero mode state
only appears in the triplet sector for a ferromagnetic 
interchain interaction.
The translation of these results to the context of the spin-1 chain is
straightforward: we just need to consider only the triplet sector. 
We find that in the Haldane gap phase ($m_t>0$) we have localized
Majorana fermion modes at the edge, but, in contrast, 
the dimerized
phase ($m_t<0$) is characterized by the absence of such degrees of 
freedom. In Ref. \onlinecite{gogolin_disordered_ladder}, it was shown
that local zero modes were associated with kinks and antikinks of the
mass $m(x)$. In the present problem, with a semi-infinite system,
the edge can be seen as a mass kink $m\theta(x)$ 
 and local zero modes should thus be
induced irrespective of the sign of $m$. However, with the
semi-infinite chain, the boundary condition (\ref{eq:majorana_bc})
selects only one chiral component. The sign of the mass then
determines whether the local mode belongs to the physical chiral
component. This is the reason for the difference of physical behavior
between positive and negative mass\cite{nersesyan_pc}.  

\subsection{The uniform component of the magnetization profile}
With all these results, the physical properties of the 
edge states of the open spin-1 chain can be investigated.
We first analyse the smooth part of the magnetization
profile of the system. To this end, we consider
the uniform part ${\bf M}(x)$ of the total spin density
(${\bf S}_+(x) = {\bf S}_1(x) + {\bf S}_2(x)$)
which takes the following form in the continuum limit
using Eq. (\ref{inispindens}):
\begin{equation}
{\bf M}\left(x\right) = \sum_{a=1}^{2} \left({\bf J}_{a R}\left(x\right)
+ {\bf J}_{a L}\left(x\right)\right).
\label{uniftotcont}
\end{equation}
With the help of the identification (\ref{currmajo}),
we immediately find that the field ${\bf M}(x)$
is expressed locally in terms of the Majorana fermions
that account for the triplet excitations in the system:
\begin{equation}
M^{a}\left(x\right) = - \frac{i}{2} \epsilon^{a b c}
\; \xi^{b}_R\left(x\right) \xi^{c}_R\left(x\right) 
- \frac{i}{2} \; \epsilon^{a b c}
\xi^{b}_L\left(x\right) \xi^{c}_L\left(x\right).
\label{unifmagmajo}
\end{equation}
Using the decomposition (\ref{decomptriplet}), we write the uniform 
density ${\bf M}(x)$  in the basis of the eigenvectors
of the Hamiltonian ${\cal H}_t$ (\ref{eq:majorana_hamiltonianrenor})
\begin{eqnarray}\label{eq:spin_density_q0}
M^{a}\left(x\right) &=& - i\epsilon^{abc} \eta^b \eta^c 
\frac{m_t}{v_t} e^{-2 m_t x/v_t} 
-\sqrt{\frac {2 m_t} {v_t L}} e^{- m_t x/v_t} 
\sum_{k > 0} \epsilon^{abc} \cos\left(k x + \theta_k^{t}\right)
\left(i \xi^{b}_{k} \eta^c + H.c.\right)  \nonumber \\
&-& \frac{\epsilon^{abc}}{2L} \sum_{k,q > 0} \left(
A\left(k,q,x\right) i \xi^{b}_{k} \xi^{c \dagger}_{q} +
B\left(k,q,x\right) i \xi^{b}_{k} \xi^{c}_{q}
+ H.c.\right), 
\end{eqnarray}
with
\begin{eqnarray}
A\left(k,q,x\right) &=& \cos\left(kx+\theta^{t}_k\right)\cos
\left(q x+\theta^{t}_{q}\right) + \sin\left(kx\right) 
\sin\left(qx\right)\nonumber \\
B\left(k,q,x\right) &=& \cos\left(kx+\theta^{t}_k\right)\cos
\left(q x+\theta^{t}_{q}\right) - \sin\left(kx\right) 
\sin\left(qx\right).
\label{idendecomp}
\end{eqnarray}
The total uniform magnetization ${\bf S}_0$ is defined by
\begin{equation}
{\bf S}_0 =\int_0^\infty dx {\bf M}\left(x\right),
\end{equation}
so that we obtain
\begin{equation}
 S_0^a = -\frac i 2 \epsilon^{abc} \eta^b \eta^c 
- 2 i \epsilon^{abc} \sum_{k > 0}  
\xi_{k}^{b} \xi_{k}^{c\dagger}.
\label{totalunifmag}
\end{equation}
The first term in this 
equation describes a spin-1/2 moment since it corresponds
to the Majorana representation of 
a spin-1/2 operator.\cite{tsvelikmajo}
In particular, the result (\ref{totalunifmag}) 
implies that the electron-spin-resonance (ESR) reponse
of the cut two-leg spin ladder with a ferromagnetic interchain
coupling  decomposes into 
the bulk response and the response at a chain end. The latter one is
identical to the ESR response of an isolated spin-1/2 impurity. 
Since there is a continuity between the weak and strong 
coupling limits in this system,\cite{hida_2ch,watanabe_ladder_obc,white} 
the Majorana approach provides thus an alternative
description of the chain-end S=1/2 mode of the open spin-1 Heisenberg chain
that has been obtained within the Schwinger bosons
formalism.\cite{ng_schwinger} 

The uniform part of the magnetization profile
of the model can be also read  from the decomposition
(\ref{eq:spin_density_q0}).
For completeness, we give in
the Appendix B an alternative derivation of 
the z-component of the uniform magnetization 
profile of the cut two-leg spin ladder without using the 
Majorana fermions method.
We obtain the following result using Eq. (\ref{eq:spin_density_q0})
\begin{equation}
\langle M^a\left(x\right)\rangle
= \frac{2 m_t}{v_t}  e^{- 2 m_t x/v_t} 
\langle -\frac i 2 \epsilon^{abc} \eta^b \eta^c  \rangle, 
\label{unifmagprofile}
\end{equation}
which can be interpreted as a spin-1/2 chain-boundary excitation 
localized over a length $v_t/2 m_t$ with an amplitude $2 m_t/v_t$.
This implies that the size of the spin-1/2 edge state
diverges while its amplitude vanishes 
as the SU(2)$_2$ WZNW critical point
of the S=1 biquadratic chain is approached from the Haldane
phase, in full agreement with the DMRG analysis of  
Ref. \onlinecite{polizzi_boundary_s=1}. The Majorana fermion
description also implies
that the uniform component of the magnetization profile should not be
affected  by temperature in the absence of an applied magnetic field.   
Let us finally mention that if the triplet mass $m_t$ is negative
then the S=1/2 chain-boundary excitations disappear as 
it can be easily seen from the decomposition (\ref{eq:eigen_decomposition}).
We thus conclude that  these free S=1/2 end spins are absent   
in a ladder with  an antiferromagnetic interchain exchange $J_{\perp}
> 0$ as well as in the dimerized phase of the spin-1 biquadratic
chain. 
It is worth noting that
the absence of free spin-1/2 moments in the spontaneously
dimerized phase of a frustrated spin chain
has been shown very recently.\cite{normand_mila}

\subsection{Calculation of the NMR relaxation rate}
The NMR relaxation rate $1/T_1$ of the cut two-leg spin
ladder with a ferromagnetic interchain interaction can be 
computed by means of the Majorana approach described in the previous
sections.
For the standard two-leg spin ladder, it has been theoretically
investigated in Refs. \onlinecite{kishine_nmr,damle_ladder,ivanov_nmr}.
We shall only consider here the uniform part of the 
NMR relaxation rate and restrict for simplicity to
the contribution that identifies to the $1/T_1$
of the spin-1 Heisenberg chain in the limit of 
strong ferromagnetic interchain coupling $-J_{\perp} \gg J_{\parallel}$.

The general formula giving this NMR relaxation rate reads
as follows\cite{kishine_nmr}
\begin{equation}
\frac 1 {T_1(x)}= \frac{T}{\omega}\textrm{Im} \chi(x,\omega),
\end{equation}
where $\omega$ is the nuclear resonance frequency which
is the smallest energy scale of the problem: $\omega \ll T,m_t$.
We introduce the following susceptibility to perform the calculation
of the NMR rate $1/T_1$:
\begin{equation}
\chi(x,i\omega_n)=\int_0^\beta d\tau e^{i\omega_n \tau} \langle T_\tau
{\bf M}\left(x,\tau\right) \cdot {\bf M}\left(x,0\right) \rangle,
\end{equation}
with the analytical continuation 
$\chi(x,\omega)=\chi(x,i\omega_n)|_{i\omega_n \to
\omega+ i0}$. 
Using the decomposition (\ref{eq:spin_density_q0}) in the 
basis of the eigenvectors
of the Hamiltonian ${\cal H}_t$ (\ref{eq:majorana_hamiltonianrenor})
that describes the triplet degrees of freedom, 
the NMR relaxation rate
can be expressed as
\begin{eqnarray}
\frac 1 {T_1(x)}= \frac{6 T \pi}{\omega L^2} \sum_{k,q >  0} 
A^2\left(k,q,x\right)
\left(n_F\left(\epsilon_k^t\right) - 
n_F\left(\epsilon_q^t\right)\right)
\delta\left(\omega+\epsilon_k^t - \epsilon_q^t\right),
\label{1t1intra}
\end{eqnarray}
$n_F(\epsilon)$ being the Fermi distribution function.
The sum in Eq. (\ref{1t1intra}) can be replaced by an integral through 
the substitution $
\sum_{k>0} \to L \int_0^\infty dk/\pi$ 
and the NMR relaxation
rate simplifies, in the low-temperature limit $T\ll m_t$, as 
\begin{equation}
\frac{1}{T_1\left(x\right)} = \frac{6}{\pi v_t^2}
\int_{0}^{\infty} dk \; e^{-\epsilon_k^t/T}
\frac{\epsilon_k^t}{\sqrt{k^2
+ \frac{2m_t\omega}{v_t^2}}} 
[\cos^2\left(kx+\theta_k^t\right) + \sin^2\left(kx\right)]^2,
\end{equation}
where the frequency $\omega$ insures the convergence of the integral at $k=0$.
Using the energy dispersion (\ref{dispersion})
of massive Majorana fermions
and the identification (\ref{eq:bogoliubov_rotation}), one 
finally obtains the expression
\begin{eqnarray}
\frac 1 {T_1\left(x\right)} &=& \frac{6}{\pi v_t^2}
\int_{0}^{\infty} dk e^{-\epsilon_k^t/T}
\frac{\epsilon_k^t }{\sqrt{k^2
+ \frac{2m_t\omega}{v_t^2}}} \left[
\frac{\left(v_t k\right)^4}{\left(\left(v_t k\right)^2+ m_t^2\right)^2}
-\frac{2m_tv_t k}{m_t^2+\left(v_t k\right)^2} 
\sin\left(2kx\right) \right.  
\nonumber \\ &+&\left.  \frac{m_t^3 v_t k}{\left(\left(v_t k\right)^2
+m_t^2\right)^2} \sin
\left(4kx\right) + \frac{2m_t^2}{\left(v_t k\right)^2
+m_t^2}\left(1-\cos\left(2kx\right)\right) + 
\frac
{m_t^2\left(\left(v_tk\right)^2-m_t^2\right)}{2\left(
\left(v_t k\right)^2+m_t^2\right)^2} 
\left(1-\cos\left(4kx\right)\right)\right].
\label{nmrexpgen}
\end{eqnarray}
At the extremity of the chain ($x=0$), the NMR relaxation rate takes
a simple form in the low-temperature limit $T \ll m_t$
\begin{equation}
\frac 1 {T_1\left(x=0\right)} = 
\frac{6 m_t}{\pi v_t^2} \left[\left(\frac T m_t -1\right)
e^{-m_t/T} + \frac{m_t}{T} E_1\left(\frac{m_t}{T}\right)\right] \sim
\left(\frac T m_t \right)^2 e^{-m_t/T},
\label{nmrchainend}
\end{equation}
$E_1(x)$ being the exponential integral function.
Therefore, we conclude that the presence of the 
boundary leads to a narrowing of NMR line at
low temperature compared to the bulk system. In principle, this NMR
rate can be measured
experimentally\cite{goto_edges_nmr} by
measurements of nuclear magnetization 
recovery.\cite{notenmr}

Now, we turn to the calculation of the $x$ dependence of $1/T_1$. 
The sine terms that appear in Eq. (\ref{nmrexpgen})
can be rewritten as
\begin{equation}
I = 
-\frac{12 m_t}{\pi v_t^2} \int_0^\infty d\theta \;
e^{-m_t \cosh \theta/T} \sin
\left(\frac{2 m_t x}{v_t} \sinh \theta \right)
+ \frac{6 m_t}{\pi v_t^2} \int_0^\infty d\theta \;
\frac{e^{-m_t \cosh \theta/T}}{\cosh^2 \theta} \sin
\left(\frac{4 m_t x}{v_t} \sinh \theta \right). 
\end{equation}
In the regime $T \ll m_t$,
this expression can be approximated as
\begin{equation}
I \simeq \frac{3 m_t}{\pi v_t^2} e^{-m_t/T} \sqrt{\frac{2T}{m_t}} 
\left(\varphi\left(\frac{2x}{\xi_T}\right) 
- 2 \varphi\left(\frac{x}{\xi_T}\right)
\right),
\end{equation}
where the function $\varphi(y)$ is defined by
\begin{equation}
\varphi\left(y\right) = \sum_{n=0}^\infty \left(-1\right)^n 
\frac{n!}{\left(2n+1\right)!} y^{2n+1},
\end{equation}
and $\xi_T$ is a thermal length which writes
\begin{equation}
\xi_T = \frac{v_t}{\sqrt{8 m_t T}}.
\label{thermalength}
\end{equation}
This length scale diverges when $T \rightarrow 0$ and 
plays the role of an effective coherence length for the NMR 
relaxation rate.
Similarily, the cosine terms of Eq. (\ref{nmrexpgen})
can be rewritten in the low-temperature limit as
\begin{equation}
J \simeq 
\frac{3 m_t}{\pi v_t^2} e^{-m_t/T} 
\int_0^\infty d\theta
\frac{e^{-m_t\theta^2/2T}}{\sqrt{ \theta^2 +
\frac{2\omega}{m_t}}} 
\left[3 - 4\cos\left(\frac {2 m_t x}{v_t} \theta \right)
+ \cos\left(\frac {4 m_t x}{v_t} \theta \right)
\right].
\end{equation}
We note that, for $x \gg a$ i.e. far from 
the chain end, the low-temperature behavior of 
the  NMR relaxation reads as follows
\begin{equation}
\frac{1}{T_1\left(x \gg a\right)} \simeq 
\frac{9 m_t}{\pi v_t^2} e^{-m_t/T} 
\int_0^\infty \frac{d\theta}{\sqrt{ \theta^2 +
\frac{2\omega}{m_t}}}  \; e^{- m_t \theta^2/ 2T},
\end{equation}
which corresponds to the bulk behavior of 
the NMR relaxation rate of the spin-1 Heisenberg
chain found in Ref. \onlinecite{sagi_nmr_haldane_gap} where the Haldane 
gap identifies to the triplet mass $m_t$. 

\section{Staggered magnetization and dimerization
profiles}\label{sec:stagg-profile}
The staggered magnetization component of a 
two-leg spin ladder with a defect has been
investigated semiclassically in
Ref. \onlinecite{fukuyama_ladder_impurity}. Such semiclassical
approach has the inconvenient of breaking SU(2) rotational
symmetry. Nevertheless, it gives useful qualitative indications on the
expected magnetization profile. For the open ladder, the
boundary condition on the
bosonic fields is $\Phi_{\pm}(0)=0$. In the bulk, a semiclassical
minimization of the ground state energy implies $\langle \Phi_+
\rangle=0$ ($J_\perp>0$) and $\langle \Phi_+\rangle=\pi/2$ ($J_\perp<0$). 
Thus, we expect no staggered magnetization profile in the case of an
antiferromagnetic rung coupling, and a profile with exponential decay
far from the boundary in the case of a ferromagnetic rung coupling. 
In this section, we present an approach that has the advantage
over the semiclassical method of preserving the full rotational
symmetry. 
As is well known, the low-energy properties
of the two-leg spin ladder can be described using four decoupled
off-critical two-dimensional Ising models.\cite{shelton_spin_ladders} 
In particular, this approach allows the calculation of 
the leading asymptotics of the staggered part of the 
spin-spin correlation functions which involve non-local
operators in terms of the underlying Majorana fermions.
In this section, we shall exploit the existence of a
similar mapping for the semi-infinite two-leg spin ladder
to determine the staggered component of the magnetization
profile and the induced dimerization in the system.

\subsection{Staggered magnetization}
Let us discuss more precisely this mapping onto an effective Ising
model. It is well known that a 1D theory of massive Majorana fermions
describes the long-distance properties of 1D quantum Ising
model.\cite{luther_ising,zuber_77,schroer_ising,ogilvie_ising}
For a recent detailed review on this
correspondence, the reader may consult for instance the 
chapter 12 of the book Ref.~\onlinecite{gogolin_book}. 
In the case of a semi-infinite system, this mapping remains valid
and the boundary conditions on the Ising spins depends on 
the ones for the Majorana fermions.\cite{ghoshal_boundary_integrable}
More specifically, we shall follow here the conventions of 
Ref. \onlinecite{ghoshal_boundary_integrable}  so that if the Majorana fermions $\xi_{R,L}$
obey the boundary condition
\begin{equation}\label{eq:free_bcs}
\xi_R(0)=\xi_L(0),
\end{equation}
then the Ising model satisfies to a free boundary condition (i.e. the
boundary spin is free to fluctuate and takes the values $\pm 1$). 
On the other hand, the Ising model experiences a 
fixed boundary condition (i.e. the
boundary spin is fixed to the value $\sigma(0)=1$ for instance)
when the Majorana fields verify
\begin{equation}\label{eq:fixed_bcs}
\xi_R(0)=-\xi_L(0).
\end{equation}
The mass $m$ of these fermions is a linear measurement of the deviation 
of the temperature with respect to the critical one: $m=T_c - T$
as in Ref. \onlinecite{ghoshal_boundary_integrable} such that a positive mass corresponds
to the low-temperature phase of the Ising model.
The low-energy 
Hamiltonian (\ref{eq:majorana_hamiltonianrenor})
of the cut two-leg spin ladder 
with the boundary conditions
(\ref{eq:majorana_bc}) on 
the fermions can thus be viewed as four decoupled off-critical
1D quantum Ising models with free boundary conditions.
In particular, the localized Majorana fermionic states with 
zero energy in the triplet sector found for
a ferromagnetic interchain coupling ($J_{\perp} < 0$) in Section III
can be interpreted physically, in the Ising mapping, as 
a domain wall attached to the boundary which separates 
two domains of opposite magnetization ($m_t = T_c - T > 0$).
In the singlet sector, one has in contrast $m_s < 0$ so that 
the corresponding Ising model with free boundary conditions
is in its disordered phase. As a consequence, 
the zero-energy Majorana mode
cannot exist in that case as it can be seen from
the decomposition (\ref{decompsinglet}).

The next step of the approach is to use 
the exact 
results
\cite{bariev_correl_ising_i,ghoshal_boundary_integrable,konik_boundary_ising}
known for the semi-infinite Ising model
to determine 
the staggered part of the magnetization profile
of the cut two-leg spin ladder. To this end,
the staggered magnetization ${\bf n}_+ = {\bf n}_1 + {\bf n}_2$
of the total spin density ${\bf S}_+ = {\bf S}_1 + {\bf S}_2$
is expressed in terms of the order and disorder operators 
$\sigma_a,\mu_a$ of the different Ising models
using the bosonic description (\ref{inspinstagboso}) and the 
bosonization approach for two 
Ising models\cite{zuber_77,schroer_ising,ogilvie_ising}
\begin{eqnarray}
n_+^x &\sim&\mu_1 \sigma_2 \sigma_3 \mu_0 \nonumber \\
n_+^y &\sim& \sigma_1 \mu_2 \sigma_3 \mu_0 \nonumber \\
n_+^z &\sim& \sigma_1 \sigma_2 \mu_3 \mu_0 .
\label{stagtotising}
\end{eqnarray}
At this point,
it is worth discussing on the ground state
degeneracy of the semi-infinite two-leg spin ladder with
a ferromagnetic interchain coupling. As it was first pointed out
by Kennedy,\cite{kennedy} an exponentially low-lying triplet,
above the singlet ground state, is found in the Haldane gap
for a finite open spin-1 Heisenberg chain. In the thermodynamic
limit, the ground state is thus fourfold degenerate.
At first sight, it seems difficult
to reproduce this result starting from three decoupled
semi-infinite quantum Ising models. Indeed, in the
strong coupling limit $-J_{\perp} \gg J_{\parallel}$,
the singlet degrees of freedom are frozen and the
three Ising models for the triplet sector are all in their
ordered phases
($m_t > 0$ for $J_{\perp} < 0$) so that 
$\langle \sigma_i \rangle \ne 0$ ($i=1,2,3$).
In that case, each Ising model has a doubly degenerate ground state
which gives thus an eightfold degeneracy. However, it is important
to note that there is a redundancy in the Ising description
since the triplet Hamiltonian in Eq. (\ref{eq:majorana_hamiltonianrenor}),
the boundary 
conditions on the Majorana fermions (\ref{eq:majorana_bc}),
the Ising representation of the 
staggered magnetization (\ref{stagtotising}) are all 
invariant under the
following transformation
\begin{eqnarray}
\xi^{i}_{R,L} &\rightarrow& - \xi^{i}_{R,L} \nonumber \\
\mu_{i} &\rightarrow& \mu_{i} \nonumber \\
\sigma_{i} &\rightarrow& -\sigma_{i},
\label{gaugesym}
\end{eqnarray}
which leads to a physical fourfold ground state degeneracy as
it should be.           
Let us return to the calculation of the magnetization profile
for a ferromagnetic interchain coupling $J_{\perp} < 0$ where
$m_t > 0$ and $m_s < 0$. 
The identification (\ref{stagtotising}) shows that
the average staggered magnetization goes to zero
far from the chain end since $\langle \mu_{1,2,3}\rangle = 0$
in the case of a positive triplet mass. 
However, due to the presence of the 
boundary, a staggered magnetization can appear close
to the chain end i.e. when $x=0$. The magnetization profile
encodes the cross-over effect on the local magnetization 
as a function of the distance from the boundary. The magnetization
profile of the spin-1 chain is obtained from the one of the ladder
with ferromagnetic interchain interaction by taking the 
limit $|m_s| \to \infty$ or
equivalently $\mu_0 \to 1$.\cite{tsvelik_field,shelton_spin_ladders}

In this respect, let us first present general results
by exploiting the duality transformation on 1D quantum Ising model.
This transformation exchanges the order and disorder operators
$\sigma \leftrightarrow \mu$ but also 
the boundary conditions on Ising spins i.e. free boundary conditions 
become fixed and vice versa. Therefore, 
one obtains the following equivalences on
the different one-point functions of the model
\begin{eqnarray}
\langle \sigma\left(T>T_c\right) \rangle_{\text{free}} &=& 
\langle \mu\left(T<T_c\right)
\rangle_{\text{fixed}}=0  \nonumber\\
\langle \sigma\left(T>T_c\right) \left(x\right)\rangle_{\text{fixed}} &=& 
\langle \mu\left(T<T_c\right) \left(x\right)
\rangle_{\text{free}} = \sigma_\infty F\left(\frac{mx}{v}\right) \nonumber \\
\langle \sigma\left(T<T_c\right)\left(x\right) \rangle_{\text{fixed}}
&=& \langle \mu\left(T>T_c\right)\left(x\right)
\rangle_{\text{free}} = \sigma_\infty G\left(\frac{mx}{v}\right) \nonumber \\
\langle \sigma\left(T<T_c\right) \left(x\right)\rangle_{\text{free}}
&=& \langle \mu\left(T>T_c\right) \left(x\right)
\rangle_{\text{fixed}} = \sigma_\infty H\left(\frac{mx}{v}\right) ,
\label{dualitycons}
\end{eqnarray}
$v$ being the velocity of the underlying Majorana fermion and
$\sigma_\infty$ is the expectation value of $\sigma$ (respectively $\mu$) for
$T<T_c$ (respectively $T>T_c$). An estimate of $\sigma_\infty$ valid for
$m\ll v/a$ is
$\sigma_\infty= 2^{1/12} e^{-1/8} A^{3/2} (|m|a/v)^{1/8}$ where
$A$
is the Glaisher constant.\cite{wu} 
It is indeed obvious that 
one has $\langle \sigma(T > T_c)\rangle_{\text{free}} = 0$
for an Ising model
with $T > T_c$ and free boundary conditions.
In contrast,
one should observe that, even in the disordered phase of the model,
a non-zero magnetization
$\langle \sigma(T > T_c)\rangle_{\text{fixed}} \ne 0$ 
exists for fixed boundary conditions since
the Ising spins are polarized at
the boundary. In that case, the precise cross-over between
the boundary and bulk behaviors is described by the function
$F$.
The staggered part
of the magnetization profile of the cut two-leg spin 
ladder with a ferromagnetic interchain coupling can thus be deduced
from the correspondence (\ref{stagtotising}) and the general results
(\ref{dualitycons})
\begin{eqnarray}
\langle n_+^{x}\left(x\right) \rangle &\sim& \left(\frac{m_t^3 |m_s|
a^4}{v_t^3 v_s}\right)^{1/8} F\left(\frac{m_t x}{v_t}\right) 
H^2\left(\frac{m_t x}{v_t}\right) G\left(\frac{|m_s| x}{v_s}\right) 
\nonumber \\
\langle n_+^{y}\left(x\right) \rangle  &\sim& \left(\frac{m_t^3 |m_s|
a^4}{v_t^3 v_s}\right)^{1/8} F\left(\frac{m_t x}{v_t}\right) 
H^2\left(\frac{m_t x}{v_t}\right) G\left(\frac{|m_s| x}{v_s}\right) 
\nonumber \\
\langle n_+^{z}\left(x\right) \rangle &\sim& \left(\frac{m_t^3 |m_s|
a^4}{v_t^3 v_s}\right)^{1/8} F\left(\frac{m_t x}{v_t}\right) 
H^2\left(\frac{m_t x}{v_t}\right) G\left(\frac{|m_s| x}{v_s}\right),
\label{stagres}
\end{eqnarray}
which exhibits a full rotationally invariant 
form as it should be.
Remarkably, a staggered magnetization 
appears although
there is none for an isolated spin-1/2 Heisenberg chain. 
The expressions of $G$ and $H$ are exactly known 
and have been determined by Bariev\cite{bariev_correl_ising_i}
from a lattice description 
and later by Konik et al.\cite{konik_boundary_ising} in the continuum
case by the form factor approach.
As it is shown in the Appendix C using this latter formalism, 
the function $F$ can, in fact, be directly expressed in terms of $G$: 
\begin{equation}
F\left(x\right) = e^{- x} G\left(x\right).
\label{frelation}
\end{equation}
As a consequence, the z-component of the staggered 
magnetization for instance simplifies as
\begin{equation}
\langle n_+^{z}\left(x\right) \rangle \sim \left(\frac{m_t^3 |m_s|
a^4}{v_t^3 v_s}\right)^{1/8} e^{- m_t x/v_t}
H^2\left(\frac{m_t x}{v_t}\right) 
G\left(\frac{m_t x}{v_t}\right) G\left(\frac{|m_s| x}{v_s}\right).
\label{stagzcompfin}
\end{equation} 
In the case of the open spin-1 chain, performing the substitution
$\mu_0 \rightarrow 1$ in Eq. (\ref{stagtotising}),
we obtain in a similar way the magnetization profile:
\begin{equation}
\langle n_+^{z}\left(x\right) \rangle \sim \left(\frac{m_t
a}{v_t}\right)^{3/8} e^{- m_t x/v_t}
H^2\left(\frac{m_t x}{v_t}\right)
G\left(\frac{m_t x}{v_t}\right).
\label{stagzcompfinchain1}
\end{equation}           
The functions $G$ and $H$ that appear in these equations 
can be cast into a Fredholm determinant form (see for instance
the Appendix C) or expressed in 
terms of a solution to the Painlev{\'e} III
differential
equation.\cite{bariev_correl_ising_i,konik_boundary_ising}
Complete expressions for $G$ and $H$ can be found in the Appendix
\ref{app:painleve}.
For the sake of simplicity, we only need here the asymptotic behaviors of
these functions which read as follows in the long-distance
limit $ X = m x/v \gg 1$\cite{bariev_correl_ising_i,konik_boundary_ising}
\begin{eqnarray}
G\left(X\right) &\simeq& 1 + \frac{1}{16\sqrt{\pi}}
\frac{e^{-2 X}}{X^{3/2}} \nonumber \\
H\left(X\right) &\simeq& 1 - \frac{1}{2\sqrt{\pi}}
\frac{e^{-2 X}}{X^{1/2}} ,
\label{longlimit}
\end{eqnarray}           
whereas in the short-distance limit $ X = m x/v \ll 1$,
one has the following estimates\cite{bariev_correl_ising_i,konik_boundary_ising}
\begin{eqnarray}
G\left(X\right) &\sim& X^{-1/8} \nonumber \\
H\left(X\right) &\sim& X^{3/8}.
\label{shortlimit}
\end{eqnarray}
>From these results, we deduce the behavior of the 
staggered component of the magnetization profile 
far from the chain end:
\begin{equation} 
\langle n_+^{z}\left(x\right) \rangle \sim \left(\frac{m_t^3 |m_s|
a^4}{v_t^3 v_s}\right)^{1/8} e^{- m_t x/v_t}
=  \left(\frac{m_t^3 |m_s|
a^4}{v_t^3 v_s}\right)^{1/8} e^{- x/\xi_t},
\label{stagzlongfree}
\end{equation} 
with a similar behavior for the open spin-1 chain.
As expected, the local staggered magnetization
decays exponentially with the distance from the boundary
with a length scale that depends only on the bulk properties
and identifies to the correlation length $\xi_t = v_t/m_t$ of the model.
It is worth noting the absence of any $x$ prefactor in
front of the exponential term in Eq. (\ref{stagzlongfree})
in the long-distance limit $x \gg \xi_t$.
This suggests that the staggered magnetization for long chains
with open boundary conditions is the relevant quantity to 
extract a very precise value of the Haldane gap as it
has been done by means of the DMRG approach.\cite{white_dmrg_letter}
A similar exponential behavior is also obtained in the semiclassical
treatment\cite{fukuyama_ladder_impurity} and in a phenomenological
theory of the open spin-1 chain describing the system as a spin-1/2
coupled to one-dimensional massive bosons.\cite{sorensen_dmrg}
Comparing to Eq.~(\ref{unifmagprofile}), we notice that the staggered
magnetization has a correlation length $\xi_t$ whereas the uniform
magnetization has a correlation length $\xi_t/2$, in agreement with
the phenomenological free boson theory.\cite{sorensen_dmrg} In
contrast to the free boson theory, we find no $x^{-3/2}$ prefactor
in the uniform component of the magnetization. 
In the short distance limit $x \ll \xi_t$, one obtains 
the following power law behavior from Eq. (\ref{shortlimit})
for the two-leg ladder with a ferromagnetic interchain interaction:
\begin{equation}
\langle n_+^{z}\left(x\right)\rangle \sim \frac {m_t}{v_t} (ax)^{1/2},
\label{stagzshortfree}
\end{equation}             
whereas, in
the case of the spin-1 chain, this power
law is modified to:
\begin{equation}
\langle n_+^{z}\left(x\right)\rangle \sim \frac{m_t a}{v_t}
\left(\frac x a\right)^{5/8}.
\label{stagzshorts1f}
\end{equation}
We note that the staggered magnetization profile (\ref{stagzcompfinchain1}) 
has a vanishing
intensity and a diverging correlation length when the Haldane gap goes
to zero, in agreement with the
DMRG analysis of Ref. \onlinecite{polizzi_boundary_s=1}. 
Moreover, 
our calculation predicts that a staggered
magnetization will exist at $T=0$
in the $S^z_{tot.}=0$ sector. This has indeed been observed in a DMRG
calculation.\cite{legeza_spins} At first sight, it seems to contradict the
results of QMC simulations.\cite{miyashita} However,
these calculations are performed at finite temperature. For the
one-dimensional quantum Ising model with free boundary conditions, 
there is no long range order in $\langle \sigma \rangle$
at $T>0$ due to the thermal
nucleation of soliton excitations. Hence, we expect that as soon as
the temperature is switched on, the average magnetization in 
$S^z_{tot.}=0$ state will
vanish in agreement with what is observed in QMC calculations.                        

The magnetization profile, in the antiferromagnetic
interchain coupling case, can be investigated by 
a similar approach. For $J_{\perp} >0$, one has
now $m_t < 0$ and $m_s > 0$ so that the Ising models of
the triplet sector are in their disordered phases whereas
the Ising model of the singlet sector belongs to its
ordered phase. We thus obtain using the 
results (\ref{stagtotising},\ref{dualitycons}) that
$\langle {\bf n}_+ \rangle = {\bf 0}$ and similarily it 
can also be shown that $\langle {\bf n}_- \rangle = 
\langle {\bf n}_1 - {\bf n}_2 \rangle = {\bf 0}$.
Therefore, we conclude on the absence of S=1/2 chain-boundary
excitations and of a non-zero magnetization profile for
the cut two-leg spin ladder with an antiferromagnetic rung coupling.
This result is consistent with the fact that the ground state of 
this model is always unique whether open or 
periodic boundary conditions are used. In this respect, 
the standard two-leg spin ladder with $J_{\perp} > 0$,
in contrast to the $J_{\perp} < 0$ case, is not
equivalent to the Haldane phase characterized by
these S=1/2 chain-end excitations eventhough they share
similar properties like the presence of a spin gap,
and a non-zero string order parameter.\cite{takada_2ch_transf,white}
In fact, it has been recently pointed out that the two systems
belong to two topologically distinct classes.\cite{kim}
In particular, it has been argued that 
the S=1 spin chain and the two-leg spin ladder with $J_{\perp} > 0$
have two different types of string order that are intimately
related to the valence bond structure of the ground states.
The topological distinction
is made by counting
the number $Q_y$ of valence bond's crossing an arbitrary
vertical line. In the case of an antiferromagnetic 
spin ladder, $Q_y$ is always even whereas it is 
odd for a system weakly connected to the spin-1 chain.
Futhermore, the authors of Ref. \onlinecite{kim} 
have noticed that for open boundary conditions
ground states characterized by an odd value of $Q_y$
have spin-1/2 edge states while 
these end states disappear when $Q_y$ is even.
This is in full agreement with the results for the 
cut two-leg spin ladder obtained in this work
within the bosonization approach.

\subsection{Dimerization induced by open boundary condition}
The dimerization profile induced by
the presence of a boundary can be also computed by this 
mapping onto semi-infinite Ising models.
The dimerization operator in terms of the original lattice spins
is defined by
\begin{equation}
\epsilon_{+ n} = \left(-1\right)^{n} \sum_{p=1}^{2} 
{\bf S}_{n,p} \cdot {\bf S}_{n+1,p} .
\label{dimerlatt}
\end{equation}
The bosonized description of this operator in the continuum
limit reads as follows in terms of the bosonic fields 
$\Phi_{\pm}$ of Eq. (\ref{symantisym})
\begin{equation}
\epsilon_+ \sim \cos \Phi_+ \cos \Phi_- .
\label{dimerboso}
\end{equation}  
Using the bosonization representation of 
two Ising models,\cite{zuber_77,schroer_ising,ogilvie_ising} this operator
can then be expressed in terms of the different Ising disorder 
operators:
\begin{equation}
\epsilon_+ \sim \mu_1 \mu_2  \mu_3  \mu_0 . 
\label{dimerising}
\end{equation} 
In the bulk, the system does not experience any
dimerization pattern since the Ising models 
in the triplet sector are in their ordered phases for $J_{\perp} < 0$
so that $\langle \mu_i \rangle = 0$ ($i=1,2,3$).
However, as for the existence of a local staggered 
magnetization, the presence of the boundary induces 
a non-trivial dimerization in the system\cite{ng_schwinger,qinng,tsai}
which can be obtained from the results (\ref{dualitycons}):
\begin{equation} 
\langle \epsilon_+ \left(x\right) \rangle \sim  \left(\frac{m_t^3 |m_s|
a^4}{v_t^3 v_s}\right)^{1/8}
e^{- 3 m_t x/v_t}
G^3\left(\frac{m_t x}{v_t}\right) G\left(\frac{|m_s| x}{v_s}\right).
\label{dimerprofile}
\end{equation} 
Using the asymptotics (\ref{longlimit}, \ref{shortlimit}), 
we deduce the following estimates for the local dimerization
\begin{eqnarray}
\langle \epsilon_+ \left(x\right) \rangle &\sim&
 \left(\frac{m_t^3 |m_s|
a^4}{v_t^3 v_s}\right)^{1/8} e^{- 3 x/\xi_t}, \; x \gg \xi_t \nonumber \\
\langle \epsilon_+ \left(x\right) \rangle &\sim& 
\left(\frac x a\right)^{-1/2}, \; \; \; x \ll \xi_t .
\label{dimerasympto}
\end{eqnarray}
Note that the exponent $1/2$ is identical to the exponent that would
have been obtained in two decoupled gapless spin-1/2 chains by boundary
conformal field
theory.\cite{eggert_openchains} Physically,
this means that the edge is making the system behaves as if it was
gapless for distances shorter than the correlation length.
In the case of the spin-1 chain, 
the dimerization operator (\ref{dimerising}) simplifies
to $\epsilon_+ \sim \mu_1 \mu_2 \mu_3$ ($\mu_0 \rightarrow 1$)
so that we get
\begin{equation}
\langle \epsilon_+ \left(x\right) \rangle \sim  \left(\frac{m_t
a}{v_t}\right)^{3/8}
e^{- 3 m_t x/v_t}
G^3\left(\frac{m_t x}{v_t}\right).
\label{dimerprofilespin1}
\end{equation}              
The long-distance limit of this dimerization
has a similar form as in Eq. (\ref{dimerasympto})
and the short distance behavior is
modified to $\langle \epsilon_+ \left(x\right) \rangle \sim 
x^{-3/8}$. Again, this exponent could have been predicted from boundary
conformal field theory.\cite{affleck_log_corr} 
We also observe that this
exponent has been obtained in the DMRG study of biquadratic spin-1 chain
at the
SU(2)$_2$ WZNW critical point.\cite{tsai}  

In the antiferromagnetic interchain coupling case, 
a similar calculation can be made. The dimerization
operator (\ref{dimerising}) has again a zero 
ground-state expectation value in the bulk since
the Ising model in the singlet sector is in its ordered phase
($m_s > 0$). As seen above, the two-leg spin ladder
with an antiferromagnetic rung coupling has no magnetic S=1/2 chain-boundary
excitations but a localized Majorana state in the singlet sector
still remains as it can be deduced from the 
decomposition (\ref{eq:eigen_decomposition}) with $m_s > 0$. 
This zero mode manifests itself in the existence of a dimerization profile
which is given by 
\begin{equation}
\langle \epsilon_+ \left(x\right) \rangle \sim  \left(\frac{|m_t|^3 m_s
a^4}{v_t^3 v_s}\right)^{1/8}
e^{- m_s x/v_t}
G^3\left(\frac{|m_t| x}{v_t}\right) G\left(\frac{m_s x}{v_s}\right),
\label{dimerprofileladder}
\end{equation}
with the following asymptotics ($\xi_s = v_s/m_s$)
\begin{eqnarray}
\langle \epsilon_+ \left(x\right) \rangle &\sim&  \left(\frac{|m_t|^3 m_s
a^4}{v_t^3 v_s}\right)^{1/8}
e^{- x/\xi_s}, \; x \gg \xi_s \nonumber \\
\langle \epsilon_+ \left(x\right) \rangle &\sim&
\left(\frac x a \right)^{-1/2}, \; \; \; x \ll \xi_s .
\label{dimerasymptoladder}
\end{eqnarray}
In the case of the spontaneously 
dimerized spin-1 chain,  
the dimerization profile takes the form
\begin{equation}
\langle \epsilon_+ \left(x\right) \rangle \sim  \left(\frac{|m_t|
a}{v_t}\right)^{3/8}
G^3\left(\frac{|m_t| x}{v_t}\right).
\label{dimerprofiledimer}
\end{equation}           
Its short distance
asymptotics 
becomes thus: $\langle \epsilon_+ \left(x\right) \rangle \sim
x^{-3/8}$, whereas the long distance one 
reads as follows using Eq. (\ref{longlimit})
\begin{equation}
\langle \epsilon_+ \left(x\right) \rangle \simeq 
\epsilon_{\infty} \left(1 + \frac{3 \xi_t^{3/2}}{16 \sqrt{\pi}}
\frac{e^{- 2 x/\xi_t}}{x^{3/2}}\right),
\label{dimerlongdist}
\end{equation}
$\epsilon_{\infty}$ being the non-zero bulk dimerization.

\section{Effect of a strong external boundary magnetic
field}\label{sec:strong-extern-field} 

The effect of a strong applied magnetic field which
fixes the spins at the boundary can be investigated using
the Ising representation described in the previous section.
To this end, let us first recall the effect of a 
transverse edge magnetic field in the semi-infinite
XXZ spin-1/2 Heisenberg chain.\cite{affleck_edge_xxz}
It has been found that the system, along 
the entire XXZ critical line, renormalizes to the infinite field
fixed point where the spin at the edge is polarized.
In the bosonization language, one has an example of a c=1
boundary flow between the Dirichlet and Neumann boundary conditions.
At the SU(2) invariant point, the edge field is exactly marginal
and a line of fixed point occurs between the Dirichlet and
Neumann limiting cases.\cite{callan,affleck_edge_xxz}
In the following, we shall only consider the physical situation
where the spin at the edge is fully polarized or fixed so that 
it corresponds to the infinite field fixed point or Neumann boundary condition
on the bosonic field $\Phi_p$ associated to the spin-1/2 chain 
with index $p=1,2$:
\begin{equation}
\partial_x \Phi_p\left(0,t\right) = 0 \; \forall t,
\label{neumann}
\end{equation}
or equivalently it can be interpreted as a Dirichlet boundary
condition on the dual field $\Theta_p$:
\begin{equation}
\Theta_p\left(0,t\right) = 0 \; \forall t .
\label{neumannbis}
\end{equation}        
The value of the constant in this expression stems from 
the fact that a magnetic field along the x-axis is considered 
in the following.
The actual direction of the applied field is not important
since the model is SU(2) invariant.
The chiral fields $\Phi_{\pm R,L}$, defined by Eq. (\ref{symantisym}),
are no longer independent due to the boundary condition (\ref{neumannbis})
and satisfy now
\begin{equation}
\Phi_{\pm L}\left(0\right) =  \Phi_{\pm R}\left(0\right),
\label{neumannchir}
\end{equation}     
from which we deduce the following analytic continuation ($x \ge 0$)
\begin{equation}
\Phi_{\pm L}\left(x,t\right) =  \Phi_{\pm R}\left(-x,t\right).
\label{neumannchirfold}
\end{equation}  

The change of boundary conditions in comparison 
to the Dirichlet case (\ref{bonphipmchir}) in zero 
field has several consequences.
First of all, the commutator between the left and right 
bosonic fields is modified due to the folding condition 
(\ref{neumannchirfold}): 
$[\Phi_{\pm R}(x), \Phi_{\pm L}(y)] = + i \pi$. 
As a consequence, the low-energy Hamiltonian of the model
for a ferromagnetic interchain coupling $J_{\perp} < 0$ 
is still given by Eq. (\ref{eq:majorana_hamiltonianrenor})
but with a negative triplet mass $m_t = J_{\perp}\lambda^2/2\pi < 0$
and a positive singlet mass $m_s = - 3 J_{\perp}\lambda^2/2\pi > 0$.
Moreover, the boundary conditions on the Majorana fermions
have also changed in the Neumann case (\ref{neumannchir}).
They can be deduced as in Section II from the identifications
(\ref{fermi_fields}, \ref{majorana}) so that we obtain the following 
boundary conditions
\begin{eqnarray}
\xi^{1}_L \left(0\right) &=& - \xi^{1}_R \left(0\right) \nonumber \\
\xi^{2}_L \left(0\right) &=& \; \; \; \xi^{2}_R \left(0\right) \nonumber \\
\xi^{3}_L \left(0\right) &=& \; \; \; \xi^{3}_R \left(0\right) \nonumber \\
\xi^{0}_L \left(0\right) &=& - \xi^{0}_R \left(0\right). 
\label{majoboundneumann}
\end{eqnarray}

One can interpret these results in light of the Ising description 
presented in the previous section.
The Ising models in the triplet sector with index $2,3$
(respectively $1$) have free (respectively fixed) 
boundary conditions and belong to their disordered phases ($m_t < 0$).
The Ising model that accounts for the singlet excitations 
has fixed boundary conditions and is in its ordered phase ($m_s > 0$).
The S=1/2 chain-boundary excitations of the open spin-1 chain
have thus disappeared in the presence of a strong applied
edge magnetic field. It remains only a single localized
Majorana fermionic state $\xi^1$ with zero energy
that describes fluctuations in the $S^x =0$ triplet subspace
which is unaffected by the applied magnetic field along
the x-direction.  The Ising representations 
of the staggered magnetization (\ref{stagtotising}) and the
dimerization operator (\ref{dimerising}) have to be modified slightly 
due to the change of sign of the commutator between
the left and right components of the bosonic fields $\Phi_{\pm}$
and they are now given by
\begin{eqnarray}
n_+^x &\sim&\sigma_1 \mu_2 \mu_3 \sigma_0 \nonumber \\
n_+^y &\sim&\mu_1 \sigma_2 \mu_3 \sigma_0 \nonumber \\
n_+^z &\sim& \mu_1 \mu_2 \sigma_3 \sigma_0 \nonumber \\
\epsilon_+ &\sim& \sigma_1 \sigma_2 \sigma_3 \sigma_0 .
\label{stagtodimertisingneumann}
\end{eqnarray}            
>From these results and the identification (\ref{dualitycons}),
we thus deduce the staggered magnetization and dimerization profiles
of the two-leg spin ladder with a ferromagnetic interchain coupling
in a strong applied magnetic field along the x-axis:
\begin{equation}
\langle n_+^{x}\left(x\right) \rangle \sim  e^{- |m_t| x/v_t}
G^3\left(\frac{|m_t| x}{v_t}\right) G\left(\frac{m_s x}{v_s}\right), 
\label{stagtodimertprofileneumann}
\end{equation}            
whereas $\langle n_+^{y} \rangle = \langle n_+^{z} \rangle =
\langle \epsilon_+ \rangle = 0$.
The asymptotics of the non-zero staggered magnetization
can be extracted from Eqs. (\ref{longlimit}, \ref{shortlimit})
\begin{eqnarray}
\langle n_+^{x}\left(x\right) \rangle &\sim&
e^{- x/\xi_t}, \; x \gg \xi_t \nonumber \\
\langle n_+^{x} \left(x\right) \rangle &\sim&
x^{-1/2}, \; \; \; x \ll \xi_t .
\label{stagpolarasympto}
\end{eqnarray}                 
 The short distance
exponent is
identical to the one predicted from boundary conformal field
theory in a gapless spin-1/2 chain with a strong magnetic field at the
boundary.\cite{affleck_edge_xxz}  The same result (\ref{stagpolarasympto})
also holds in the case of the spin-1 chain, albeit with a different
short distance behavior $\langle n_+^{x} \left(x\right) \rangle \sim
x^{-3/8}$ which can be obtained from boundary conformal field
theory. We conclude that the staggered magnetization in 
a strong applied field decays in the same way 
as in Eq. (\ref{stagzlongfree}) far from the boundary but is enhanced
in the vicinity of the chain end in comparison to
the behavior (\ref{stagzshortfree}) in zero field. These results are
in agreement with  QMC
simulations of the spin-1 Heisenberg chain with free and fixed
boundary conditions.\cite{miyashita}

A similar calculation can be made in the case of 
an antiferromagnetic interchain interaction.
The only
difference is that we must make the following
substitution $T-T_c \to T_c-T$. For a strong applied 
field along the x-direction, we get now
\begin{equation}
\langle n_+^{x}\left(x\right) \rangle \sim e^{-2m_t x/v_t - |m_s| x/v_s} 
G^3\left(\frac{m_t x}{v_t}\right) G\left(\frac{|m_s| x}{v_s}\right),
\label{profilestaggfixedladder}
\end{equation}
and 
$\langle n_+^y \rangle = \langle n_+^z \rangle =0$.
However, there is now a non-zero staggered relative 
magnetization $\langle {\bf n}_- = {\bf n}_1 - {\bf n}_2\rangle$
in the $J_{\perp} > 0$ case which can be determined using
the Ising representation of this operator:
\begin{eqnarray}
n_-^x &\sim&\mu_1 \sigma_2 \sigma_3 \mu_0 \nonumber \\
n_-^y &\sim&\sigma_1 \mu_2 \sigma_3 \mu_0 \nonumber \\
n_-^z &\sim&\sigma_1 \sigma_2 \mu_3 \mu_0, 
\label{stagrelaisingneumann}
\end{eqnarray}
so that 
\begin{equation}
\langle n_-^{y,z}\left(x\right) \rangle \sim e^{-m_t x/v_t}
G^2\left(\frac{m_t x}{v_t}\right) 
H\left(\frac{m_t x}{v_t}\right) 
H\left(\frac{|m_s| x}{v_s}\right),
\label{profilerelastaggfixedladder}
\end{equation}
and $\langle n_-^x \rangle = 0$.
Finally, the dimerization profile in the $J_{\perp} > 0$ case
reads as follows
\begin{equation}
\langle \epsilon_+\left(x\right) \rangle \sim e^{-|m_s| x/v_s}
G\left(\frac{m_t x}{v_t}\right) H^2\left(\frac{m_t x}{v_t}\right)
G\left(\frac{|m_s| x}{v_s}\right),
\label{dimerprofilefixedladder}
\end{equation}
so that we obtain the following asymptotics respectively
in the long and short distance limits
\begin{eqnarray}
\langle \epsilon_+\left(x\right) \rangle &\sim& e^{- x/\xi_s}
\nonumber \\
\langle \epsilon_+\left(x\right) \rangle &\sim& x^{1/2} .
\end{eqnarray}

We close this section by discussing the case of the spontaneously
dimerized spin-1 chain. 
The Ising representations of staggered and dimerization 
fields are now given by Eq. (\ref{stagtodimertisingneumann}) 
with $\sigma_0  \rightarrow 1$.
The staggered magnetization and dimerization
profiles are 
\begin{eqnarray}
\langle n_+^{x}\left(x\right) \rangle &\sim& 
e^{- 2x/\xi_t}
G^3\left(\frac{x}{\xi_t}\right)  \nonumber \\
\langle \epsilon_+\left(x\right) \rangle &\sim&
G\left(\frac{x}{\xi_t}\right) H^2\left(\frac{x}{\xi_t}\right).
\label{stagdimerprofiledimer}
\end{eqnarray}    
In the short distance limit, we obtain the following power law 
behaviors
\begin{eqnarray}
\langle n_+^{x}\left(x\right) \rangle &\sim& x^{-3/8} \nonumber \\
\langle \epsilon_+\left(x\right) \rangle &\sim& x^{5/8}, 
\end{eqnarray}            
whereas for long distances we have
\begin{eqnarray}
\langle n_+^{x}\left(x\right) \rangle &\sim& e^{-2 x/\xi_t} 
\nonumber \\
\langle \epsilon_+\left(x\right) \rangle &\simeq& 
\epsilon_{\infty} \left(1 - \frac{\xi_t^{1/2}}{\sqrt{\pi}}
\frac{e^{- 2 x/\xi_t}}{x^{1/2}}\right) .
\end{eqnarray}
Therefore, we observe that the 
dimerization reaches here its bulk
expectation value from below in contrast to the 
free boundary case (\ref{dimerlongdist}).

\section{Concluding remarks}\label{sec:concluding-remarks}

In this paper, we have investigated the nature of the
chain-boundary excitations of the cut two-leg spin ladder
and the open spin-1 chain by means of the bosonization
method.
The crucial point of the analysis is 
the mapping\cite{shelton_spin_ladders,tsvelik_field} 
of the low-energy Hamiltonian of these systems onto
free massive Majorana fermions (or 
equivalently decoupled non-critical quantum Ising
models) with suitable boundary conditions.
In particular, the exact 
results\cite{bariev_correl_ising_i,ghoshal_boundary_integrable,konik_boundary_ising}
of the semi-infinite one-dimensional quantum Ising model allow 
the determination of the low-energy properties of the cut
two-leg spin ladder such as, for instance, the magnetization and
dimerization profiles. For a ferromagnetic rung coupling ($J_{\perp} < 0$),
the system is characterized by the presence of fractional
spin-1/2 edge states which, in the limit $J_{\perp}/J_{\parallel} 
\rightarrow -\infty$, identify to the well known S=1/2 chain-end
degrees of freedom of the open spin-1 chain.
In this respect, the approach, presented in this paper, 
provides an alternative derivation of the existence of these
edge states first predicted 
theoretically within the VBS model\cite{affleck_klt_short} 
and the Schwinger boson 
mean-field analysis.\cite{ng_schwinger}
In the case of an antiferromagnetic interchain interaction
$J_{\perp} > 0$, no S=1/2 chain-end excitations are found but
a non-magnetic localized Majorana fermion zero mode
is still present and leads to the formation of a non-zero
dimerization profile in the system.

The magnetization and dimerization profiles, derived in this paper,
should be confronted to numerical simulations of the cut
two-leg spin ladder or the open biquadratic spin-1 chain in
the vicinity of the SU(2)$_2$ WZNW critical point.
Due to the semi-infinite geometry considered here, 
our results would best be compared with DMRG calculations of a finite
spin-1 chain with a spin-1/2 attached to one of the extremities to
cancel one of the edge states.
\cite{white_dmrg_letter,legeza_spins}
At this point, it is worth noting that all the calculations were
done at zero temperature.
Our results could in principle be extended to 
finite temperature using the
thermal form factor techniques derived for
the one-dimensional quantum Ising model.\cite{leclair_qising_thermal}
Unfortunately,
it is not an easy task within this formalism to obtain explicit
expressions for $\langle \sigma(x)\rangle$ for fixed boundary
conditions.  This makes difficult any direct comparison to
QMC simulations.\cite{miyashita,alet_doped_s=1}
However, one can argue by finite size
scaling arguments that the correlation functions should not be 
strongly affected by
a finite temperature as long as $m\gg T$. This can be checked by
an explicit computation of the two-point correlation
function.\cite{sachdev_ising} We also stress that at
finite temperature and for free boundary conditions, one 
has $\langle \sigma(x)\rangle=0$ in the quantum Ising model.
This implies the
absence of any staggered magnetization profile in the $S=0$ state 
and also of a non-zero string order parameter at finite
temperature in
agreement with the QMC simulations.\cite{miyashita}

Regarding perspectives,
the approach, presented in this work, could be 
applied to other one-dimensional gapful systems.
The effects of an uniaxial single-ion anisotropy 
$D_z \sum_i (S_i^z)^2$ on the magnetic properties 
of the open spin-1 chain can be investigated.
Since the different species of Majorana fermions do
not interact, we expect that spin-1/2 edge states excitations should
still be observed, in agreement with the QMC
results.\cite{yamamoto_anistropys1_qmc} The calculation of
magnetization profiles with our method should not pose any
difficulty. The approach could also be generalized 
to study the effect of a weak bond or magnetic impurities
in a spin-1 chain\cite{sorensenaffleck} as it will be discussed
in a separate publication. Another interesting situation is
the nature of the edge states of two open spin-1/2 Heisenberg chains
coupled by a biquadratic interchain interaction.\cite{nersesyan_biquadratic}
Due to the extended O(4) symmetry of the model, we expect to
find two spin-1/2 excitations at the edge of the system.
A more challenging problem is the generalization of our
approach to $S\ge 3/2$ Heisenberg chains. According to
Ref.~\onlinecite{ng_schwinger}, it is expected that edge states
with fractionalized spin $S^{\prime}$ exist in the spin-S chain 
with $S^{\prime} = S/2$ (respectively $S^{\prime} = (S-1/2)/2$)
for integer (respectively half-odd-integer) spins.
This conjecture, based on the large-N limit of SU(N) quantum
antiferromagnets and strong-coupling expansion, has been 
verified by a DMRG analysis for $S=3/2,2$.\cite{qinng}
A generalization of the approach presented here in 
the S=1 case is to describe spin-S Heisenberg chain as 
perturbed $SU(2)_{2S}=U(1)\otimes Z_{2S}$ 
WZNW models.\cite{affleck_strongcoupl,cabra_spin_s}
For $S$ half-odd integer, only the
parafermion sector $Z_{2S}$ is gapped. We should thus expect the edge
spin excitations to be generated by the bound states of the massive
$Z_{2S}$ theory on the half-line. For $S$ integer, edge excitations
should be induced by the boundary bound states of the perturbed WZNW model. 
A similar problematic should also be considered for the related
problem of $n$-leg spin-1/2 ladders. 
Finally, an interesting question would be the study of interactions
between edge states in chains of finite length\cite{batista} using the
Majorana fermions description. 

\begin{acknowledgments}
The authors would like to thank T. Jolicoeur, D. Allen,
P. Azaria, M. Bocquet, A. A. Nersesyan, M. Saito, and
Y. Suzumura for valuable discussions. E. O. thanks Nagoya University
for kind hospitality and support during his stay at the Physics Department 
where this project was initiated. 
\end{acknowledgments}

\appendix
\section{Bosonization approach of the open S=1/2 Heisenberg
chain}\label{app:boson-open-s=12} 

In this appendix, we describe the bosonization approach 
of the spin-1/2 Heisenberg chain with open 
boundary conditions.\cite{eggert_openchains,wong,ng_qin,hikihara}
This enables us to fix the conventions 
that will be used in this paper and also to discuss some
subtleties related  
to the presence of open boundary conditions.
To this end, we consider the repulsive Hubbard model
at half-filling with open boundary conditions described
by the Hamiltonian
\begin{equation}  
{\cal H}_U = -t\sum_{i=1}^{N-1}\left(c_{i \sigma}^{\dagger} c_{i+1 \sigma}
+ H. c. \right) + U \sum_{i=1}^{N} n_{i\uparrow} n_{i\downarrow}, 
\label{hubbardopen}
\end{equation}
where $c_{i\sigma}$ is the electronic annihilation operator
of spin index $\sigma = \uparrow, \downarrow$ at site $i$ 
($1 \le i \le N$) and $n_{i\sigma} = c_{i \sigma}^{\dagger} c_{i \sigma}$
stands for the occupation number of electron with spin index $\sigma$.
The summation over repeated
greek symbols is assumed in the following and the 
hopping term $t$ is positive.
In this model, it is well known that a 
charge gap $m_c$ exists for any
positive value of the interaction $U$ and in the low-energy 
limit ($E \ll m_c$) only the spin excitations remain and describe
the universal scaling properties of the spin-1/2 Heisenberg chain.
In this way, we shall derive the continous description of the 
spin density of the S=1/2 Heisenberg chain with open boundary conditions
starting from the electronic model (\ref{hubbardopen}).
An alternative approach as described 
in Refs. \onlinecite{eggert_openchains,ng_qin,hikihara}
is to consider the spin-1/2 XXZ Heisenberg chain with open
boundary conditions and the use of the Jordan-Wigner transformation.
Since in this work we shall only consider SU(2) invariant
interactions, it is more appropriate to start from the 
Hubbard model (\ref{hubbardopen}).
The open boundary conditions are taken into account
by introducing two fictious
sites $0$ and $N+1$ in Eq. (\ref{hubbardopen}) and
by imposing vanishing boundary conditions on the fermion
operators: $c_0 = c_{N+1} = 0$.\cite{eggert_openchains,fabrizio_open_electron_gas}
The low-energy properties of the model can then be 
determined by applying   the bosonization method\cite{gogolin_book}
with suitable boundary conditions on the 
bosonic fields.\cite{eggert_openchains,wong,fabrizio_open_electron_gas,mattson}

\subsection{Non-interacting case}
In the low-energy limit, the continuum version of the non-interacting
part of the Hamiltonian (\ref{hubbardopen})
can be derived by expressing the lattice fermions $c_{n\sigma}$ in terms of 
left and right moving spinful fermionic fields $\Psi_{L,R \sigma} (x)$: 
\begin{equation}
\frac{c_{n \sigma}}{\sqrt{a}} 
\sim i^{x/a} \; \Psi_{R \sigma} \left(x\right)
+ \left(-i\right)^{x/a} \; \Psi_{L \sigma}\left(x\right),
\label{contferdes}
\end{equation}
with $x= n a$, $a$ being the lattice spacing.
The resulting boundary conditions on the fermionic fields
of Eq. (\ref{contferdes}) are thus 
\begin{eqnarray}
\Psi_{L\sigma}\left(0\right) &=& - \Psi_{R\sigma}\left(0\right) \nonumber \\
\Psi_{L\sigma}\left(L\right) &=& -
\left(-1\right)^{L/a} \Psi_{R\sigma}\left(L\right), 
\label{boundfermxy}
\end{eqnarray}
with $L = (N+1)a$.
The left and right excitations are no longer independent
due to the presence of these boundaries.

The next step of the approach is the introduction 
of right and left moving bosonic fields $\Phi_{R,L\sigma}$
through 
\begin{eqnarray} 
\Psi_{R\sigma} &=& \frac{\kappa_{\sigma} 
e^{i \pi \tau_{\sigma}/4}}{\sqrt{2\pi a}} \; 
e^{-i \Phi_{R\sigma}} \nonumber \\
\Psi_{L\sigma} &=& \frac{\kappa_{\sigma} 
e^{i \pi \tau_{\sigma}/4}}{\sqrt{2\pi a}} \; 
e^{i \Phi_{L\sigma}},
\label{abelboso}
\end{eqnarray}
where $\kappa_{\sigma}$ are Klein factors that obey the anticommutation
relations $\{\kappa_{\sigma}, \kappa_{\sigma \prime} \} = \delta_{\sigma,
\sigma \prime}$ to ensure the anticommutation between the 
fermion fields of different spin index.
In Eq. (\ref{abelboso}), we have also introduced some phase factors
with $\tau_{\uparrow} = 1$ and $\tau_{\downarrow} = -1$ for
later convenience.
The boundary conditions on the chiral bosonic fields 
are then obtained from Eq. (\ref{boundfermxy}):
\begin{eqnarray}
\Phi_{L\sigma}\left(0\right) &=& - \Phi_{R\sigma}\left(0\right) + \pi \nonumber \\
\Phi_{L \sigma} \left(L\right) &=& - \Phi_{R\sigma} \left(L\right) + 
\pi \left(\frac{L}{a} -1\right) + 2 q_{\sigma} \pi,
\label{boundbosmxy}
\end{eqnarray}
$q_{\sigma}$ being an integer.
In our conventions, the total bosonic field $\Phi_{\sigma}$
with spin index $\sigma$ and its dual $\Theta_{\sigma}$
are related to the chiral components $\Phi_{R,L\sigma}$ through 
\begin{eqnarray} 
\Phi_{\sigma} &=& \frac{1}{2} \left(\Phi_{R \sigma} 
+ \Phi_{L \sigma} \right) \nonumber \\
\Theta_{\sigma} &=& \frac{1}{2} \left(\Phi_{L \sigma} 
- \Phi_{R \sigma} \right),
\label{phitheta}
\end{eqnarray}
so that Eq. (\ref{boundbosmxy}) imposes
Dirichlet boundary conditions on the bosonic field: 
\begin{eqnarray}
\Phi_{\sigma}\left(0\right) &=& \frac{\pi}{2} \nonumber \\
\Phi_{\sigma}\left(L\right) &=&
\frac{\pi}{2}\left(\frac{L}{a} -1\right) + q_{\sigma} \pi . 
\label{boundbostotmxy}
\end{eqnarray}

The low-energy dynamics of the non-interacting Hamiltonian
${\cal H}_0$ of the original model (\ref{hubbardopen})
is thus described by two independent free massless boson Hamiltonian
with the boundary conditions (\ref{boundbostotmxy}):
\begin{equation}
{\cal H}_0 = \frac{v_F}{2\pi}\sum_{\sigma=\uparrow, \downarrow}
\int_{0}^{L}dx \; 
\left(\left(\partial_x \Phi_{\sigma}\right)^2 
+ \left(\pi \Pi_{\sigma}\right)^2\right),
\label{freeboshamappen}
\end{equation}
$v_F$ being the Fermi velocity 
and $\Pi_{\sigma}$ is the momentum operator conjugate
to $\Phi_{\sigma}$. 
The mode decomposition of the bosonic field $\Phi_{\sigma}$ compatible
with these boundary conditions reads as follows
\begin{equation}
\Phi_{\sigma}\left(x,t\right) = \frac{\pi}{2} + \left(\frac{\pi}{2} 
\left(\frac{L}{a} - 2\right) + 
\sqrt{\pi}\; {\tilde \pi}_{0 \sigma} \right) \frac{x}{L} 
+ \sum_{n=1}^{\infty}
\frac{\sin\left(k_n x\right)}{\sqrt{n}} 
\left(\alpha_{n \sigma} e^{-i k_n v_F t} + H.c. \right),
\label{bosmode}
\end{equation}
where $k_n = n \pi/L$, $\alpha_{n \sigma}$ is the boson annihilation
operator obeying $[\alpha_{n \sigma}, \alpha_{m \sigma \prime}^{\dagger}] 
= \delta_{n,m} \delta_{\sigma, \sigma\prime}$
and the zero mode operator ${\tilde \pi}_{0\sigma}$ has a discrete
spectrum $\sqrt{\pi} q_{\sigma}$. The mode decomposition 
of the momentum 
operator $\Pi_{\sigma} = \partial_t \Phi_{\sigma}/\pi v_F$ conjugate
to the bosonic field can thus be deduced from Eq. (\ref{bosmode})
\begin{equation}
\Pi_{\sigma}\left(x,t\right) = \sum_{n=1}^{\infty} \frac{i\sqrt{n}}{L} 
\sin\left(k_n x\right)
\left(-\alpha_{n \sigma} e^{-i k_n v_F t} 
+ \alpha_{n \sigma}^{\dagger}e^{i k_n v_F t}  \right).
\label{piopmode}
\end{equation}
In particular, one can check that the mode 
decompositions (\ref{bosmode}, \ref{piopmode})  
satisfy the canonical commutation relation: 
\begin{equation}
[\Phi_{\sigma}\left(x,t\right), \Pi_{\sigma\prime}\left(y,t\right)] 
= i \delta_{\sigma, \sigma\prime} \delta_L(x-y),
\label{canorel}
\end{equation}
$\delta_L(x-y)$ being the delta function at finite size:
$\delta_L(x) = \sum_n e^{i k_n x}/2 L$.
The dual field $\Theta_{\sigma}$ satisfies
$\partial_x \Theta_{\sigma} = \pi \Pi_{\sigma}$ 
and $\partial_t \Theta_{\sigma} = v_F\partial_x\Phi_{\sigma}$ so that 
one obtains 
the following mode expansion:
\begin{equation}
\Theta_{\sigma} \left(x,t\right) = \sqrt{\pi}\; {\tilde \phi}_{0 \sigma}
+ \left(\frac{\pi}{2}
\left(\frac{L}{a} - 2\right) +
\sqrt{\pi}\; {\tilde \pi}_{0 \sigma} \right) \frac{v_F t}{L}   
+ i \sum_{n=1}^{\infty} \frac{\cos\left(k_n x\right)}{\sqrt{n}} 
\left(\alpha_{n \sigma} e^{-i k_n v_F t} 
- \alpha_{n \sigma}^{\dagger}e^{i k_n v_F t} \right),
\label{dualmode}
\end{equation}
where the zero mode coordinate ${\tilde \phi}_{0\sigma}$ 
is conjugate to ${\tilde \pi}_{0\sigma}$: 
$[{\tilde \phi}_{0\sigma}, {\tilde \pi}_{0\sigma^{'}}]
= i\delta_{\sigma, \sigma^{'}}$ and it
is not
fixed by the boundary conditions (\ref{boundbostotmxy}). 
Finally,
the mode decompositions of the chiral bosonic
fields $\Phi_{R,L\sigma}$ can be determined by 
the identification (\ref{phitheta}):
\begin{eqnarray}
\Phi_{L \sigma} &=&
\frac{\pi}{2} + \left(\frac{\pi}{2}
\left(\frac{L}{a} - 2\right) +
\sqrt{\pi}\; {\tilde \pi}_{0 \sigma} \right) \frac{x + v_F t}{L} +
\sqrt{\pi}\; {\tilde \phi}_{0 \sigma}
+ \sum_{n=1}^{\infty}  \frac{i}{\sqrt{n}}
\left(\alpha_{n \sigma} e^{-i k_n\left(x + v_F t\right)} - 
\alpha_{n \sigma}^{\dagger} e^{i k_n\left(x + v_F t\right)} \right) \nonumber \\
\Phi_{R \sigma} &=& 
\frac{\pi}{2} + \left(\frac{\pi}{2}
\left(\frac{L}{a} - 2\right) +
\sqrt{\pi}\; {\tilde \pi}_{0 \sigma} \right) \frac{x - v_F t}{L} -
\sqrt{\pi}\; {\tilde \phi}_{0 \sigma}  
- \sum_{n=1}^{\infty}  \frac{i}{\sqrt{n}}
\left(\alpha_{n\sigma} e^{i k_n\left(x - v_F t\right)} -
\alpha_{n\sigma}^{\dagger} e^{-i k_n\left(x - v_F t\right)} \right)
\label{chirbosmodesigma}
\end{eqnarray}
In addition, one can show that these chiral fields satisfy the following
commutation relations when $L \gg a$
\begin{eqnarray}
\left[\Phi_{L \sigma} \left(x,t\right), \Phi_{L \sigma \prime} \left(y,t\right)
\right] &=& -i \pi \delta_{\sigma, \sigma \prime}
\ \mathrm{sgn}\left(x - y\right) \nonumber \\
\left[\Phi_{R \sigma} \left(x,t\right), \Phi_{R \sigma \prime} \left(y,t\right)
\right] &=& i \pi \delta_{\sigma, \sigma \prime}
\ \mathrm{sgn} \left(x - y\right) \nonumber \\
\left[\Phi_{R \sigma} \left(x,t\right), \Phi_{L \sigma \prime} \left(y,t\right)
\right] &=& 0  \ \ \ \ \ \ \ \ \ \ \ \ \mathrm{if} \ x = y = 0 \nonumber \\
        &=& - 2 i \pi \delta_{\sigma, \sigma \prime} \ \mathrm{if} \ x = y = L\nonumber \\ 
        &=& - i \pi \delta_{\sigma, \sigma \prime} \ \ \ \mathrm{if} \ 
        0< x, y < L,
\label{chircomm}
\end{eqnarray}
${\rm sgn}(x)$ being the sign function.

\subsection{Effective spin density}
The next step of the approach is the introduction of the bosonic fields 
that describe the charge and spin degrees of freedom:
\begin{eqnarray}
\Phi_{c R,L} &=& \frac{\Phi_{R,L\uparrow}
+ \Phi_{R,L\downarrow}}{\sqrt{2}}  \nonumber \\
\Phi_{s R,L} &=& \frac{\Phi_{R,L\uparrow}
- \Phi_{R,L\downarrow}}{\sqrt{2}}.
\label{spinchargebasis}
\end{eqnarray}   
This basis as well as the commutation relations (\ref{chircomm})
allow us to express the Hamiltonian
(\ref{freeboshamappen}) in terms of two commuting
gapless spin and charge contributions: 
\begin{equation}
{\cal H}_0 = \frac{v_F}{2\pi}
\int_{0}^{L}dx \;
\left(\left(\partial_x \Phi_{c}\right)^2
+ \left(\pi \Pi_{c}\right)^2\right) + 
\frac{v_F}{2\pi}
\int_{0}^{L}dx \;
\left(\left(\partial_x \Phi_{s}\right)^2
+ \left(\pi \Pi_{s}\right)^2\right).
\label{spinchargeham0}
\end{equation}
As well known, a weak Hubbard interaction preserves this
famous spin-charge separation and opens at half-filling a mass gap $m_c$
for the charge degrees of freedom. 
In the spin sector, the effect of the interaction is exhausted 
by a renormalization of the spin velocity and by the existence
of a marginal irrelevant contribution in the Hamiltonian.
In particular, the interaction does not renormalize the 
bosonic field $\Phi_{s}$ since it is protected by the 
underlying SU(2) symmetry of the model.
Neglecting the logarithmic corrections introduced by the marginal irrelevant
term, the low-energy ($E \ll m_c$) Hamiltonian that describes
the universal properties of the S=1/2 Heisenberg chain is simply
\begin{equation}
{\cal H}_s = \frac{v_s}{2\pi} \int_{0}^{L}dx \;
\left(\left(\partial_x \Phi_{s}\right)^2
+ \left(\pi \Pi_{s}\right)^2\right),
\label{spinhameff}
\end{equation}  
$v_s$ being the velocity of the spin collective mode.
The boundary conditions of the bosonic field $\Phi_s$ 
can be obtained from Eqs. (\ref{boundbostotmxy}, \ref{spinchargebasis}):
\begin{eqnarray}
\Phi_{s}\left(0\right) &=& 0 \nonumber \\
\Phi_{s}\left(L\right) &=&
\frac{q \pi}{\sqrt{2}},
\label{boundbosspin}
\end{eqnarray}          
$q$ being an integer.
Similarily, the mode decompositions of the chiral bosonic
fields $\Phi_{sR,L}$ read as follows with help 
of Eq. (\ref{chirbosmodesigma})
\begin{eqnarray}
\Phi_{s L}\left(x,t\right) &=&
\sqrt{\pi}\; {\tilde \pi}_{0 s} \; \frac{x + v_s t}{L} +
\sqrt{\pi}\; {\tilde \phi}_{0 s}
+ \sum_{n=1}^{\infty}  \frac{i}{\sqrt{n}}
\left(\alpha_{n s} e^{-i k_n\left(x + v_s t\right)} -
\alpha_{n s}^{\dagger} e^{i k_n\left(x + v_s t\right)} \right) 
\nonumber \\
\Phi_{s R}\left(x,t\right) &=&
\sqrt{\pi}\; {\tilde \pi}_{0 s} \; \frac{x - v_s t}{L} -
\sqrt{\pi}\; {\tilde \phi}_{0 s}
- \sum_{n=1}^{\infty}  \frac{i}{\sqrt{n}}
\left(\alpha_{n s} e^{i k_n\left(x - v_s t\right)} -
\alpha_{n s}^{\dagger} e^{-i k_n\left(x - v_s t\right)} \right),
\label{chirmodespin}
\end{eqnarray}
with $[{\tilde \phi}_{0 s}, {\tilde \pi}_{0 s}] = i$ 
and $[\alpha_{n s}, \alpha_{m s}^{\dagger}] = \delta_{n,m}$.
Moreover, we deduce from Eq. (\ref{chircomm}) that
these fields obey the commutation relations
\begin{eqnarray}
\left[\Phi_{s L} \left(x,t\right), \Phi_{s L} \left(y,t\right)
\right] &=& -i \pi 
\mathrm{sgn}\left(x - y\right) \nonumber \\
\left[\Phi_{s R} \left(x,t\right), \Phi_{s R} \left(y,t\right)
\right] &=& i \pi 
\mathrm{sgn} \left(x - y\right) \nonumber \\
\left[\Phi_{s R} \left(x,t\right), \Phi_{s L} \left(y,t\right)
\right] &=& 0  \ \ \ \ \  \ \mathrm{if} \ x = y = 0 \nonumber \\
        &=& - 2 i \pi \ \mathrm{if} \ x = y = L\nonumber \\
        &=& - i \pi \ \ \ \mathrm{if} \  0< x, y < L .
\label{chircommspin}
\end{eqnarray}
At this point, it is important to note that the last
commutator in Eq. (\ref{chircommspin}) has the opposite
sign of the prescription made 
in Refs. \onlinecite{shelton_spin_ladders,gogolin_book}.
The actual value of its sign stems from the fact that 
we are considering open boundary conditions in the lattice
system which identify to the Dirichlet boundary 
conditions (\ref{boundbosspin}) on the  bosonic field $\Phi_s$.
In fact, one can derive the value of the commutator 
$[\Phi_{s R}, \Phi_{s L}]$ by a different method.
The boundary condition at $x=0$ on the chiral bosonic spin fields
is
\begin{equation}
\Phi_{s L}\left(0,t\right) = - \Phi_{s R}\left(0,t\right) \; \forall t .
\label{bondchirspin0}
\end{equation}
Since $\Phi_{s L}\left(x,t\right) = \Phi_{s L}\left( x + v_s t\right)$
and  $\Phi_{s R}\left(x,t\right) = \Phi_{s R}\left(v_s t - x\right)$,
one thus obtains the folding condition ($x \ge 0$):
\begin{equation}
\Phi_{s L}\left(x,t\right) = - \Phi_{s R}\left(-x,t\right), 
\label{folding}
\end{equation}          
which is satisfied by the mode decompositions (\ref{chirmodespin}).
Moreover, the commutator $[\Phi_{s R}(x,t), \Phi_{s R}(y,t)]$
is fixed by the requirement that $\Phi_s$ and $\Pi_s$ are
canonical conjugate operators so that by using
the folding condition (\ref{folding}) we deduce:
\begin{eqnarray}
\left[\Phi_{s R} \left(x,t\right), \Phi_{s L} \left(y,t\right)\right] &=&
- \left[\Phi_{s R} \left(x,t\right), \Phi_{s R} \left(-y,t\right)\right]
\nonumber \\
&=& - i \pi \ \mathrm{sgn}\left(x + y\right) = - i \pi .
\label{chircombis}
\end{eqnarray} 
It turns out that the sign of this commutator is important
for the investigation of the S=1/2 chain-boundary excitations 
of the open two-leg spin ladder as described in Sections II and III
of this work.

With all these results at hands, it is straightforward
to derive the continuum description of the spin density
starting from the lattice spin operator ${\bf S}_i$:
\begin{equation}
S_i^{a} = \frac{1}{2} \ c_{i\alpha}^{\dagger} {\sigma}^{a}_{\alpha \beta}
c_{i\beta},
\label{spinopelatt}
\end{equation}
${\sigma}^{a} (a=x,y,z)$ being the Pauli matrices.
Using the decomposition (\ref{contferdes}),
the spin density separates
into a uniform and staggered parts in the continuum limit:
\begin{equation}
{\bf S}\left(x\right) = {\bf J}_{sR}\left(x\right)
+ {\bf J}_{sL}\left(x\right) 
+ \left(-1\right)^{x/a} {\bf n}_s\left(x\right),
\label{spindecomp}
\end{equation}
with the identification
\begin{eqnarray} 
J_{sR,L}^{a} &=& \frac{1}{2} \ \Psi_{R,L\alpha}^{\dagger}
{\sigma}^{a}_{\alpha \beta} \Psi_{R,L\beta} \nonumber \\ 
n_s^{a} &=& \frac{1}{2}\left(\Psi_{L\alpha}^{\dagger}
{\sigma}^{a}_{\alpha \beta} \Psi_{R\beta}
+ \Psi_{R\alpha}^{\dagger}
{\sigma}^{a}_{\alpha \beta} \Psi_{L\beta}\right).
\label{spindecompfer}
\end{eqnarray}  
The bosonized description of the spin density can then
be obtained with help of the bosonization formula (\ref{abelboso}),
the commutation relations (\ref{chircomm}),
and the canonical transformation (\ref{spinchargebasis}). The 
resulting expressions for the uniform part read as follows
\begin{eqnarray} 
J_{sL}^{z} &=& - \frac{1}{2\pi\sqrt{2}} \ \partial_x \Phi_{s L} \nonumber \\
J_{sR}^{z} &=& - \frac{1}{2\pi\sqrt{2}} \ \partial_x \Phi_{s R} \nonumber \\
J_{sR}^{\dagger} &=& - \frac{i\kappa_{\uparrow} \kappa_{\downarrow}}{2\pi a}
\ e^{i \sqrt{2} \Phi_{s R}} \nonumber \\ 
J_{sL}^{\dagger} &=& - \frac{i\kappa_{\uparrow} \kappa_{\downarrow}}{2\pi a}
\ e^{-i \sqrt{2} \Phi_{s L}},
\label{spinuniboso}
\end{eqnarray}      
whereas the staggered part is given by
\begin{eqnarray}
n_s^{x} &=& - \frac{\lambda i\kappa_{\uparrow} \kappa_{\downarrow}}{\pi a} 
\ \cos\left(\sqrt{2} \ \Theta_s\right) \nonumber \\
n_s^{y} &=& \frac{\lambda i\kappa_{\uparrow} \kappa_{\downarrow}}{\pi a} 
\ \sin\left(\sqrt{2}\ \Theta_s\right) \nonumber \\
n_s^{z} &=& - \frac{\lambda}{\pi a} \ \sin\left(\sqrt{2}\ \Phi_s\right),
\label{spinstagboso}
\end{eqnarray}     
$\lambda$ being a constant stemming from the charge degrees of
freedom that have been integrated out in the 
low-energy regime $E \ll m_c$. The product 
$\kappa_{\uparrow} \kappa_{\downarrow}$ has no dynamic and in
this work we use the 
prescription $\kappa_{\uparrow} \kappa_{\downarrow} =i$ for
simplicity. Finally, as it can be checked
from the correspondence (\ref{spinuniboso}), the left-moving contribution 
of the uniform part of the spin density (\ref{spindecomp}) 
obeys the following operator product expansion (with 
a similar result for the right-moving term)
\begin{equation}
J_{s L}^{a}\left(z\right) J_{s L}^{b}\left(w\right) \sim
\frac{\delta_{a,b}}{8\pi^2\left(z-w\right)^2} + 
\frac{i \epsilon^{a b c} J_{s L}^{b}\left(w\right)}{2\pi\left(z-w\right)},
\label{su21ope}
\end{equation}
with $z=v_s \tau + i x$. The uniform left spin density ${\bf J}_{sL}$
identifies to the SU(2)$_1$ Kac-Moody currents which 
are the generators of the 
conformal field theory associated to the criticality 
of the spin-1/2 Heisenberg chain (see for instance the book\cite{gogolin_book}
for a review). 

\section{Alternative derivation of the uniform magnetization
profile}\label{app:altern-deriv-profile}
In this Appendix, we derive the z-component of 
the uniform magnetization profile of the
cut two-leg spin ladder with
a ferromagnetic rung coupling without 
using the Majorana fermion formalism. To this end, 
we return to the complex fermion Hamiltonian ${\cal H}_+$ (\ref{ham2legdirac})
with $J_{\perp} < 0$. 
The z-part of the total magnetization density is given by:
$M^z = :\psi^\dagger_{+R}\psi_{+R}+\psi^\dagger_{+L}\psi_{+L}:$. 
Using the boundary condition (\ref{eq:continuation_fermi}) and the
Hamiltonian (\ref{ham2legdirac}) with $m=-J_{\perp}/2\pi > 0$, 
we obtain the following mode decomposition for the
fermionic fields $\psi_{+ R,L}$:
\begin{eqnarray}
\psi_{+R}(x)& = & \sqrt{\frac m v} e^{-mx/v} a_0 + \frac 1 {\sqrt{2L}}
\sum_k (f(k,x) a_{k,+} +f^*(k,x) a_{k,-}) \nonumber \\
\psi_{+L}(x)& = & \sqrt{\frac m v} e^{-mx/v} a_0 + \frac 1 {\sqrt{2L}}
\sum_k (f^*(k,x) a_{k,+} +f(k,x) a_{k,-}) ,  \\
\end{eqnarray}
\noindent where $f(k,x)=\cos(kx+\theta_k)+i\sin(kx)$, $\theta_k$ being
defined by Eq. (\ref{eq:bogoliubov_rotation}).
The Hamiltonian ${\cal H}_+$ (\ref{ham2legdirac}) can be
expressed in terms of the $a_{k,\pm}$ fermionic modes
\begin{equation}
{\cal H}_+=\sum_k \epsilon(k) (a^\dagger_{k,+} a_{k,+} -
a^\dagger_{k,-} a_{k,-}),
\label{hplusappen}
\end{equation}
with the energy dispersion $\epsilon(k) = \sqrt{v^2 k^2 + m^2}$.
The uniform magnetization profile along the z-axis is then 
given by
\begin{equation}
\langle M^z(x)\rangle =\frac 1 {L} \sum_{k>0} [\cos^2(kx +\theta_k) + \sin^2(kx) -1] (\langle
a^\dagger_{k,-} a_{k,-} + a^\dagger_{k,+} a_{k,+} \rangle) +
\frac{2m}{v} e^{-2mx/v} \langle a^\dagger_0 a_0 \rangle .
\end{equation}
Using the expressions (\ref{eq:bogoliubov_rotation}), noting that
$n_F(\epsilon(k))+n_F(-\epsilon(k))=1$ and 
$\sum_{k > 0} \to L \int_0^\infty dk/\pi$ in the large $L$ limit,
the uniform magnetization simplifies as
\begin{eqnarray}
\langle M^z(x) \rangle &= & \frac{2m}{v} e^{-2mx/v} \langle a^\dagger_0
a_0\rangle\nonumber \\ &  -&  \int_0^\infty \frac {dk}{\pi}
\left[\frac{m^2}{(vk)^2+m^2}\cos(2kx) +\frac{mvk}{(vk)^2+m^2}\sin(2kx)\right].
\end{eqnarray}
Performing the integrals, we finally find the following result
\begin{equation}
\langle M^z(x)\rangle = \frac{2m}{v}  e^{-2mx/v} [ \langle a^\dagger_0 a_0
\rangle -1/2]. 
\end{equation}
This profile is identical to the one obtained by the Majorana fermions
calculation (\ref{unifmagprofile}). 

If we apply a uniform magnetic field along the $z$-axis, the
Hamiltonian ${\cal H}_+$  (\ref{hplusappen}) is modified as follows
\begin{equation}
{\cal H}_+ =\sum_{k,r=\pm} (r \epsilon(k)-h):a^\dagger_{k,r} a_{k,r}: -h
(a^\dagger_0 a_0-1/2), 
\end{equation}
whereas the Hamiltonian ${\cal H}_-$ in Eq. (\ref{hcont}) is not 
affected by the magnetic field. 
The resulting free energy per unit length is then given by
\begin{equation}
f_s=-\frac 1 \beta \int_0^\infty \frac{dk}{\pi} \sum_{r=\pm} \ln
(1+e^{-\beta(\epsilon(k)-rh)}) -\frac 1 \beta \ln (2\cosh(\beta h/2))=f_s^{\text{bulk}}+f_s^{\text{edge}}.
\end{equation}
>From $f_s^{\text{bulk}}$, we recover the usual susceptibility and
specific heat of the spin-1 chain. We see that $f_s^{\text{edge}}$ is
the free energy of an isolated spin-1/2. Thus, for $h\ll \beta$ and
$h\ll m$, the thermodynamic properties of the system are identical
to those of an isolated spin-1/2. This result is in agreement with the
QMC simulations of long chains\cite{miyashita} and DMRG calculations
of effective interaction of edge states in long chains\cite{batista}. 

\section{Calculation of the function $F(x)$}\label{app:calc-funct-fx}

In this Appendix, we compute the one-point function 
of the disorder operator in the low-temperature phase
of the semi-infinite one-dimensional quantum Ising model
with free boundary conditions.
To this end, the form factor approach to correlation 
functions\cite{saleur} will be 
used as in Ref. \onlinecite{konik_boundary_ising}
for the calculation of the magnetization one-point function.

Let us first recall some
results
on the form factors of the 
bulk quantum Ising model.\cite{berg_formfactors_ising,yurov} 
The excited states of this model
are created by acting on the ground state with fermion creation
operators $A^\dagger$:
\begin{equation}
|\theta_1 \ldots \theta_n \rangle = A^\dagger(\theta_1) \ldots
A^\dagger(\theta_n) |0\rangle,
\end{equation}
where the $\theta_i$'s are the usual rapidity variables parametrizing
momentum and energy as $p(\theta_i) = m \sinh \theta_i$, 
$e(\theta_i) = m \cosh \theta_i$, m being the fermion's mass 
($m > 0$ for $T < T_c$) 
and its velocity has been set to unity here for simplicity.
The creation $A^\dagger$ and annihilation $A$ operators
satisfy the fermionic anticommutation relation normalized 
as follows
\begin{equation}
\{ A\left(\theta_1\right), A^\dagger \left(\theta_2\right) \}
= 2 \pi \delta\left(\theta_1 - \theta_2\right) .
\end{equation}  
For $T<T_c$, the form factors of the order operator $\sigma$ are:
\begin{equation}\label{eq:sigma_ff}
\langle 0 | \sigma(0) |\theta_1 \ldots \theta_{2n}\rangle = i^n \prod_{1\le
i < j \le 2n} \tanh\left(\frac{\theta_i - \theta_j} 2 \right),
\end{equation}
whereas the form factors with an odd number of rapidities are zero. 
They are normalized such that the conformal limit of the 
spin-spin correlation function is
\begin{equation}
\langle \sigma\left(r\right) \sigma\left(0\right) \rangle =
\frac{{\cal F}^2}{r^{1/4}}, 
\end{equation}
with $r =\sqrt{x^2 + \tau^2}$ and ${\cal F} = 2^{-1/12} e^{1/8}
A^{-3/2} m^{-1/8}$, $A$ being the Glaisher constant. 
In the low-temperature phase, 
the form factors of the disorder operator are given by
\begin{equation}\label{eq:mu_ff}
\langle 0 | \mu(0) |\theta_1 \ldots \theta_{2n+1}\rangle =i^n \prod_{1\le
i < j \le 2n+1} \tanh\left(\frac{\theta_i - \theta_j} 2 \right),
\end{equation}
and those with an even number of rapidities are zero.
For $T>T_c$, the roles of $\sigma$ and $\mu$ are interchanged.

With these results, one can extend the
method of Ref. \onlinecite{konik_boundary_ising} to calculate
the one-point function of the disorder operator in 
the low-temperature phase of
the semi-infinite Ising model with free boundary conditions.
The free boundary condition on the Majorana fermions 
($\xi_L(0) = \xi_R(0)$) is interpreted as a boundary state
$|B \rangle$ which encodes all informations about the boundary
condition.\cite{ghoshal_boundary_integrable}
In this approach, the Hilbert space of the theory is the same
as in the bulk so that the one-point function can be extracted
through
\begin{equation}
\langle \mu \left(x\right) \rangle = \sum_n \langle 0 | 
\mu\left(0\right) |n \rangle
\langle n | B \rangle e^{-x E_n},
\label{muexpan}
\end{equation}
$|n>$ being a complete set of states of the bulk Hilbert space.
The boundary state corresponding to the Ising model at $T < T_c$
on the half line with free boundary 
conditions is\cite{ghoshal_boundary_integrable}  
\begin{equation}\label{eq:boundary_state_free}
|B \rangle = \left(1+A^\dagger\left(0\right)\right) 
\exp \left[ \int_{0}^{\infty} \frac{d\theta}{2\pi}
\hat{R}(\theta) A^\dagger(-\theta) A^\dagger(\theta) \right] |0 \rangle,
\end{equation}
with $\hat{R}(\theta)=-i \coth(\theta/2)$.
This boundary state 
contains a zero-momentum one-particle state which corresponds
to a domain wall, attached to the boundary, that separates two 
domains of opposite magnetization ($T < T_c$).
Such term contributes to the expectation value (\ref{muexpan})
while it does not enter the calculation of the one-point
function of the order operator. 
Expanding the exponential in Eq. (\ref{eq:boundary_state_free}),
we get the following expression using the fact that the
form factors of $\mu$ are non-zero only for an odd
number of rapidities:
\begin{equation}\label{eq:ff_exp_mu_free}
\langle \mu (x) \rangle =\sum_{n=0}^\infty \frac 1 {n!}
\int_{0}^\infty \frac{d\theta_1}{2\pi} \ldots
\int_{0}^\infty \frac{d\theta_n}{2\pi} \langle 0 | \mu(0) | 0;
-\theta_1, \theta_1; \ldots ; -\theta_n, \theta_n \rangle
\hat{R}(\theta_1) \ldots \hat{R}(\theta_n) e^{-mx [1+2\cosh(\theta_1)
+\ldots +2\cosh(\theta_n)]}, 
\end{equation}
from which we deduce the identity
\begin{equation}
\langle \mu(x) \rangle = e^{-mx} A(mx).
\label{muArela}
\end{equation}
The next step of the approach is to use the form factor of 
$\mu$ (\ref{eq:mu_ff}) to derive
an expression for $A(mx)$. First of all, one has
\begin{eqnarray}
\langle 0 |\mu(0)| 0; \theta_1, -\theta_1 ;\ldots; \theta_n,-\theta_n
\rangle= i^n  \prod_{i=1}^n \tanh^2 \frac {\theta_i} 2 \tanh
\theta_i \prod_{i<j} 
\tanh^2\left(\frac{\theta_i -\theta_j} 2 \right) 
\tanh^2\left(\frac{\theta_i +\theta_j} 2 \right),
\end{eqnarray}
so that, using the expression of $\hat{R}(\theta)$,
we obtain:
\begin{eqnarray}
A(mx)=\sum_{n=0}^\infty \frac{1}{n!}\int_{0}^\infty
\frac{d\theta_1}{2\pi}\ldots \int_{0}^\infty
\frac{d\theta_n}{2\pi} \left( \prod_{i=1}^n \tanh \frac {\theta_i} 2
\tanh \theta_i\right)\; {\rm det} W(\theta_i,\theta_j)
\;e^{- 2mx\sum_{k=1}^{n} \cosh(\theta_k)},
\end{eqnarray}
with
\begin{equation}
 W(\theta_i,\theta_j)=\frac{2\sqrt{\cosh \theta_i \cosh
\theta_j}}{\cosh \theta_i + \cosh \theta_j} .
\end{equation}
Following Ref. \onlinecite{konik_boundary_ising}, we introduce
the quantity
\begin{equation}
V(\theta_i,\theta_j,mx)= 
\frac{\sqrt{\cosh \theta_i -1} \sqrt{\cosh \theta_j -1}}{
\cosh \theta_i+ \cosh \theta_j}
\; \; 
e^{-mx\left(\cosh \theta_i + 
\cosh \theta_j\right)} .
\end{equation}
The function $A(mx)$ can then be expressed as a Fredholm determinant:
\begin{eqnarray}
A\left(mx\right) &=& \sum_{n=0}^\infty \frac{1}{n!}\int_{-\infty}^\infty
\frac{d\theta_1}{2\pi}\ldots \int_{-\infty}^\infty   
\frac{d\theta_n}{2\pi} \; {\rm det} V\left(\theta_i,\theta_j,mx\right) 
\nonumber \\
&=& {\rm Det}\left(1 + \frac{V}{2\pi} \right).
\label{fredholmA}
\end{eqnarray}
Using the results obtained in Ref. \onlinecite{konik_boundary_ising},
$A(mx)$, given by Eq. (\ref{fredholmA}), coincides with
the Fredholm determinant representation of the one-point function
of the Ising magnetization for $T<T_c$ with fixed boundary conditions
i.e. $G(mx)$ in our notations (see in particular Eq. (\ref{dualitycons})).
Therefore, from Eq. (\ref{muArela}),
we finally deduce the following relation:
\begin{equation}
F(mx)=e^{-mx} G(mx) .
\end{equation}

\section{Expression of the functions $G(x)$ and $H(x)$ in terms of
solutions of the Painlev\'e III equation}\label{app:painleve}
According to
Bariev,\cite{bariev_correl_ising_i} the
functions $G$ and $H$, describing
the cross-over effect on the local magnetization 
of the semi-infinite Ising model at $T<T_c$  
with free and fixed boundary conditions, can be expressed in terms of a 
solution
$\eta(\theta)$  of the
Painlev\'e III differential equation. This latter equation reads
as follows
\begin{equation}\label{eq:painleve3}
\frac 1 \eta \frac{d^2\eta}{d\theta^2}=\left(\frac 1 \eta
\frac{d\eta}{d \theta}\right)^2 -\frac{1}{\theta \eta}\frac{d\eta}{d \theta}
+\eta^2  -\frac 1 {\eta^2}.
\end{equation}
The boundary conditions on $\eta$ are:
\begin{eqnarray}
\eta(\theta)&\sim& -\theta\left[\ln \frac \theta 4 +\gamma_E\right] (\theta
\to 0) \nonumber \\
\eta(\theta)&\sim& 1-\frac{K_0(2\theta)}{2\pi} (\theta \to \infty), 
\end{eqnarray}
$\gamma_E$ being the Euler's constant.
The functions $G$ and $H$ that are the building blocs 
of the staggered magnetization and dimerization profiles
are related to the solution $\eta(\theta)$ by
\begin{eqnarray}
G(y) &=&\eta^{-1/4}(y) \exp\left[\int_y^\infty d\theta \left\{ \frac
\theta 8 \eta^{-2}(\theta) \left( (1-\eta^2(\theta))^2
-\left(\frac{d\eta}{d\theta}\right)^2 \right)-\frac 1 2
(1-\eta(\theta))\right\} \right] \nonumber \\
H(y) &=& \eta^{1/4}(y) \exp\left[\int_y^\infty d\theta \left\{ \frac
\theta 8 \eta^{-2}(\theta) \left( (1-\eta^2(\theta))^2
-\left(\frac{d\eta}{d\theta}\right)^2 \right)-\frac 1 2
(\eta^{-1}(\theta)-1)\right\} \right].
\label{GHsolpainleve}
\end{eqnarray}

There is an interesting connection between the Painlev\'e III
differential 
equation and the two-dimensional sinh-Gordon equation. 
Indeed, the relation is obtained by 
considering $\eta(\theta)=e^{-\chi(\theta)}$ so that
the differential equation (\ref{eq:painleve3}) takes the form
\begin{equation}
\frac{d^2\chi}{d\theta^2} +\frac 1 \theta \frac{d\chi}{d\theta}=2\sinh2\chi .
\end{equation}
The functions G and H in Eq. (\ref{GHsolpainleve}) can then 
be expressed in terms of $\chi$
\begin{eqnarray}
G(y) &=& e^{\chi(y)/4}\exp\left[\int_y^\infty d\theta \left\{\frac \theta 8
\left[4 \sinh^2 \chi - \left(\frac{d\chi}{d\theta}\right)^2\right]
-\frac 1 2 (1-e^{-\chi(\theta)})\right\} \right] \nonumber \\
H(y) &=& e^{-\chi(y)/4}\exp\left[\int_y^\infty d\theta \left\{\frac \theta 8
\left[4 \sinh^2 \chi - \left(\frac{d\chi}{d\theta}\right)^2\right]
-\frac 1 2 (e^{\chi(\theta)}-1)\right\} \right] .
\end{eqnarray}


\begin{thebibliography}{101}
\bibitem{haldane_gap}
F.~D.~M. Haldane,
Phys. Rev. Lett. \textbf{50},
1153 (1983).
\bibitem{glarum}
S. H. Glarum, S. Geschwind, K. M. Lee, M. L. Kaplan,
and J. Michel,
Phys. Rev. Lett. \textbf{67}, 1614 (1991).
The spin-1/2 end states have, in fact, first been observed in
the NENP compound doped with magnetic ions:
M. Hagiwara, K. Katsumata, I. Affleck, B. I. Halperin, and J. P. Renard,
{\sl ibid.} {\bf 65}, 3181 (1990).
\bibitem{tedoldi}
F. Tedoldi, R. Santachiara, and M. Horvati{\'c},
Phys. Rev. Lett. \textbf{83},
412 (1999). 
\bibitem{uchiyama_s1_imp} 
Y. Uchiyama, Y. Sasago, I. Tsukada, K. Uchinokura,
A. Zheludev, T. Hayashi, N. Miura, and
P. B{\"o}ni,
Phys. Rev. Lett. \textbf{83},
632 (1999).
\bibitem{martins}
G. B. Martins, M. Laukamp, J. Riera, and
E. Dagotto, Phys. Rev. Lett. \textbf{78},
3563 (1997);
M. Laukamp, G. B. Martins, C. Gazza, A. L. Malvezzi,
E. Dagotto, P. M. Hansen, A. C. L{\'o}pez, and J. Riera,
Phys. Rev. B \textbf{57},
10 755 (1998).
\bibitem{azuma_zinc_doping}
M. Azuma, Y. Fujishiro, M. Takano, M. Nohara,
and H. Takagi,
Phys. Rev. B \textbf{55}, R8658 (1997).
\bibitem{fujiwara_srcuo_imp}
N. Fujiwara, H. Yasuoka, Y. Fujishiro, M. Azuma, and 
M. Takano,
Phys. Rev. Lett. \textbf{80},
604 (1998).
\bibitem{ohsugi_srcuo_imp}
S. Ohsugi, Y. Tokunaga, K. Ishida, Y. Kitaoka,
M. Azuma, Y. Fujishiro, and M. Takano,
Phys. Rev. B \textbf{60},
4181 (1999).
\bibitem{larkin_2leg_imp}
M. I. Larkin, Y. Fudamoto, I. M. Gat, A. Kinkhabwala,
K. M. Kojima, G. M. Luke, J. Merrin, B. Nachumi, 
Y. J. Uemura, M. Azuma, T. Saito, and M. Takano, 
Phys. Rev. Lett. 
\textbf{85}, 1982 (2000).
\bibitem{deguchi_cuhpcl_imp}
H. Deguchi, S. Sumoto, S. Takagi, M. Mito, T. Kawae, 
K. Takeda, H. Nojiri, T. Sakon, and M. Motokawa,
J. Phys. Soc. Jpn. \textbf{67},
3707 (1998).
\bibitem{hase_cugeo3_zn}
M. Hase, I. Terasaki, Y. Sasago, K. Uchinokura, 
and H. Obara,
Phys. Rev. Lett. \textbf{71},
4059 (1993);
L. P. Regnault, J. P. Renard, G. Dhalenne, and
A. Revcholevschi,
Europhys. Lett. \textbf{32},
579 (1995);
S. B. Oseroff, S-W. Cheong, B. Aktas, M. F. Hundley, Z.
Fisk, and L. W. Rupp, Jr, 
Phys. Rev. Lett. \textbf{74},
1450 (1995).
\bibitem{affleck_klt_short}
I. Affleck, T. Kennedy, E. H. Lieb, and H.
Tasaki,
Phys. Rev. Lett. \textbf{59},
799 (1987);
Comm. Math. Phys. \textbf{115},
477 (1988).
\bibitem{kennedy}
T. Kennedy, J. Phys.: Condens. Matter \textbf{2},
5737 (1990).
\bibitem{kennedy_z2z2_haldane}
T. Kennedy and H. Tasaki,
Phys. Rev. B \textbf{45},
304 (1992).
\bibitem{nijs_dof}
M. P.  M. den Nijs and 
K. Rommelse,
Phys. Rev. B \textbf{40},
4709 (1989).
\bibitem{kim}
E. H. Kim, G. F{\'a}th, J. S{\'o}lyom,
and D. J. Scalapino,
Phys. Rev. B \textbf{62}, 14 965 (2000).
\bibitem{ng_schwinger}
T. K. Ng, Phys. Rev. B \textbf{45}, 8181 (1992);
\textbf{47}, 11 575 (1993);
\textbf{50},
555 (1994).
\bibitem{miyashita}
S. Miyashita and S. Yamamoto,
Phys. Rev. B \textbf{48},
913 (1993);
9528 (1993).
\bibitem{white_dmrg_letter}
S. R. White,
Phys. Rev. Lett. \textbf{69},
2863 (1992);
Phys. Rev. B \textbf{48},
10 345 (1993);
S.~R. White and D. A. Huse,
{\sl ibid.} \textbf{48},
3844 (1993).
\bibitem{sorensen_dmrg}
E. S. S{\/o}rensen and I. Affleck,
Phys. Rev. B \textbf{49},
15 771 (1994). 
\bibitem{qinng} 
S. Qin, T. K. Ng, and Z. B. Su,
Phys. Rev. B \textbf{52},
12 844 (1995).
\bibitem{batista}
C.~D. Batista, K. Hallberg and A.~A. Aligia,
Phys. Rev. B \textbf{58},
9248 (1998); Phys. Rev. B  \textbf{60}, 12553 (1999). 
\bibitem{legeza_spins}
{\"O}. Legeza and J. S{\'o}lyom,
Phys. Rev. B \textbf{59},
3606 (1999).
\bibitem{alet_doped_s=1}
F. Alet and E. S. S{\/o}rensen,
Phys. Rev. B \textbf{62},
14 116 (2000).
\bibitem{polizzi_boundary_s=1}
E. Polizzi, F. Mila, and E. S. S{\/o}rensen,
Phys. Rev. B \textbf{58},
2407 (1998).
\bibitem{takhtajan_spin_s}
L. Takhtajan, Phys. Lett. A \textbf{87},
479 (1982);
J. Babujian,
{\sl ibid.} \textbf{90},
479 (1982).
\bibitem{affleck_strongcoupl}
I. Affleck and F. D. M. Haldane,
Phys. Rev. B \textbf{36},
5291 (1987).
\bibitem{gogolin_book}
For a review, see for instance: A. O. Gogolin, A. A. Nersesyan,
and A. M. Tsvelik,
\emph{Bosonization and 
Strongly Correlated Systems}, (Cambridge University
  Press, Cambridge, 1998).
\bibitem{hida_2ch}
K. Hida, J. Phys. Soc. Jpn. \textbf{60}, 1347 (1991);
\textbf{64}, 4896 (1995).
\bibitem{watanabe_ladder_obc}
H. Watanabe,
Phys. Rev. B \textbf{50},
13 442 (1994).
\bibitem{white}
S. R. White,
Phys. Rev. B \textbf{53},
52 (1996).
\bibitem{shelton_spin_ladders}
D. G. Shelton, A. A. Nersesyan,
and A. M. Tsvelik, Phys. Rev. B
\textbf{53}, 8521 (1996).
\bibitem{gogolin_disordered_ladder} 
A. O. Gogolin, A. A. Nersesyan, A. M. Tsvelik,
and L. Yu,
Nucl. Phys. B \textbf{540},
705 (1999).
\bibitem{tsai}
S. W. Tsai and J. B. Marston,
Phys. Rev. B \textbf{62},
5546 (2000).
\bibitem{eggert_openchains}
S. Eggert and I. Affleck,
Phys. Rev. B \textbf{46},
10 866 (1992);
Phys. Rev. Lett. \textbf{75},
934 (1995).
\bibitem{wong}
E. Wong and I. Affleck,
Nucl. Phys. B \textbf{417},
403 (1994).
\bibitem{fabrizio_open_electron_gas}
M. Fabrizio and A. O. Gogolin,
Phys. Rev. B \textbf{51},
17 827 (1995).
\bibitem{ng_qin}
T. K. Ng, S. Qin, and Z. B. Su,
Phys. Rev. B \textbf{54},
9854 (1996).
\bibitem{mattson}
A. E. Mattson, S. Eggert, and H. Johannesson,
Phys. Rev. B \textbf{56},
15 615 (1997).
\bibitem{hikihara}
T. Hikihara and A. Furusaki,
Phys. Rev. B \textbf{58},
R583 (1998).
\bibitem{schulz_spins}
H. J. Schulz,
Phys. Rev. B \textbf{34},
6372 (1986).
\bibitem{zamolodchikov_fateev}
A. B. Zamolodchikov
and V. A. Fateev,
Yad. Fiz. \textbf{43}, 1031 (1986)
[Sov. J. Nucl. Phys. {\bf 43}, 657 (1986)].
\bibitem{allen}
D. Allen and D. S{\'e}n{\'e}chal,
Phys. Rev. B \textbf{55},
299 (1997).
\bibitem{tsvelik_field}
A. M. Tsvelik,
Phys. Rev. B \textbf{42},
10 499 (1990).
\bibitem{lopatin}
A. V. Lopatin and V. M. Yakovenko,
cond-mat/0106516.
\bibitem{balents_random_dirac}
L. Balents and M. P. A. Fisher,
Phys. Rev. B \textbf{56},
12 970 (1997).
\bibitem{shelton_disorder}
D. G. Shelton and A. M. Tsvelik,
Phys. Rev. B \textbf{57},
14 242 (1998).
\bibitem{gogolinsupra}
A. O. Gogolin,
Phys. Rev. B \textbf{54},
16 063 (1996).
\bibitem{bassi}
Z. S. Bassi and 
A. LeClair,
Phys. Rev. B \textbf{60},
615 (1999).
\bibitem{kitaev}
A. Kitaev,
cond-mat/001440.
\bibitem{levitov_qbits}
L. S. Levitov,
T. P. Orlando, 
J. B Majer, and J. E. Mooij, 
cond-mat/0108266.
\bibitem{nersesyan_pc}
A. A. Nersesyan,
private communication. 
\bibitem{tsvelikmajo}
A. M. Tsvelik,
Phys. Rev. Lett. \textbf{69},
2142 (1992).
\bibitem{normand_mila}
B. Normand and F. Mila,
cond-mat/0109095.
\bibitem{kishine_nmr}
J. Kishine and H. Fukuyama,
J. Phys. Soc. Jpn. \textbf{66},
26 (1997).
\bibitem{damle_ladder}
K. Damle and S. Sachdev,
Phys. Rev. B \textbf{57},
8307 (1998).
\bibitem{ivanov_nmr}
D. A. Ivanov and P. A. Lee,
Phys. Rev. B \textbf{59},
4803 (1999).
\bibitem{goto_edges_nmr} 
T. Goto, S. Satoh, Y. Matsumura,
and M. Hagiwara,
Phys. Rev. B \textbf{55},
2709 (1997);
T. Goto, S. Satoh, K. Kitamura, N. Fujiwara,
J. P. Renard, and M. W. Weisel,
J. Mag. Mag. Mat. 
\textbf{177-181}, 663 (1998);
S. Satoh, T. Goto, M. Yamashita, and T. Ohishi,
{\sl ibid.} \textbf{177-181},
689 (1998).
\bibitem{notenmr} 
In Refs. \onlinecite{goto_edges_nmr},
the quantity $T_1^{-1}(x=0)$ is called $\tau_e^{-1}$.
\bibitem{sagi_nmr_haldane_gap}
J. Sagi and I. Affleck,
Phys. Rev. B \textbf{53},
9188 (1996).
\bibitem{fukuyama_ladder_impurity} 
H. Fukuyama, N. Nagaosa, 
M. Saito, and T. Tanimoto,
J. Phys. Soc. Jpn. \textbf{65},
2377 (1996).
\bibitem{luther_ising}
A. Luther and I. Peschel,
Phys. Rev. B \textbf{12},
3906 (1975).
\bibitem{zuber_77}
J. B. Zuber and C. Itzykson,
Phys. Rev. D \textbf{15},
2875 (1977).
\bibitem{schroer_ising}
B. Schroer and  T. T. Truong,
Nucl. Phys. B \textbf{144},
80 (1978).
\bibitem{ogilvie_ising}
M. Ogilvie, Ann. Phys. \textbf{136},
273 (1981).
\bibitem{ghoshal_boundary_integrable}
S. Ghoshal and A. B. Zamolodchikov,
Int. J. Mod. Phys. A \textbf{9},
3841 (1994).
\bibitem{bariev_correl_ising_i}
R. Z. Bariev,
Teor. Mat. Fiz. \textbf{40},
95 (1979);
\textbf{42},
262 (1980);
\textbf{77},
127 (1988).
\bibitem{konik_boundary_ising} 
R. Konik, A. LeClair, and 
G. Mussardo,
Int. J. Mod. Phys. A \textbf{11},
2765 (1996).
\bibitem{wu} 
T. T. Wu, B. M. McCoy, C. A. Tracy, and E. Barouch, 
Phys. Rev. B \textbf{13}, 316 (1976).
\bibitem{takada_2ch_transf}
S. Takada and H. Watanabe,
J. Phys. Soc. Jpn. \textbf{61},
39 (1992).
\bibitem{affleck_log_corr}
I. Affleck, D. Gepner, H. J. Schulz, and T. Ziman, 
J. Phys. A: Math. Gen. \textbf{22},
511 (1989).
\bibitem{affleck_edge_xxz}
I. Affleck, J. Phys. A: Math. Gen. \textbf{31}, 2761 (1998).
\bibitem{callan}
C. G. Callan, I. R. Klebanov, A. W. W. Ludwig, and J. M.
Maldacena,
Nucl. Phys. B \textbf{422}, 417 (1994).
\bibitem{leclair_qising_thermal} 
A. Leclair, F. Lesage, S. Sachdev, 
and H. Saleur,
Nucl. Phys. B \textbf{482},
579 (1996).
\bibitem{sachdev_ising}
S. Sachdev,
Nucl. Phys. B \textbf{464}
576  (1996).
\bibitem{yamamoto_anistropys1_qmc}
S. Yamamoto and S. Miyashita,
Phys. Rev. B \textbf{50},
6277 (1994).
\bibitem{sorensenaffleck}
E. S. Sorensen and I. Affleck,
Phys. Rev. B {\bf 51}, 16 115 (1995). 
\bibitem{nersesyan_biquadratic}
A. A. Nersesyan and A. M. Tsvelik,
Phys. Rev. Lett. \textbf{78},
3939 (1997), {\sl ibid.} \textbf{79}, 1171(E).
\bibitem{cabra_spin_s} 
D. C. Cabra, P. Pujol, and C. von Reichenbach,
Phys. Rev. B \textbf{58},
65 (1998).
\bibitem{saleur} 
For a recent review, see 
H. Saleur, in New Theoretical Approaches to Strongly Correlated 
Systems, edited by A. M. Tsvelik, NATO Science Series, 
(Kluwer Academic Publishers, Dordrecht, 2001).
\bibitem{berg_formfactors_ising}
B. Berg, M. Karowski, and P. Wiesz, Phys. Rev. D {\bf 19},  2477  (1979).
\bibitem{yurov} 
V. P. Yurov and Al. B. Zamolodchikov, Int. J. Mod. Phys. A {\bf 16},
3419 (1991); J. L. Cardy and G. Mussardo, Nucl. Phys. B {\bf 340}, 
387 (1990).
\end{thebibliography}
\end{document}